\documentclass[sigconf]{acmart}

\AtBeginDocument{%
  \providecommand\BibTeX{{%
    \normalfont B\kern-0.5em{\scshape i\kern-0.25em b}\kern-0.8em\TeX}}}

\copyrightyear{2024}
\acmYear{2024}
\setcopyright{acmlicensed}\acmConference[CHI '24]{Proceedings of the CHI Conference on Human Factors in Computing Systems}{May 11--16, 2024}{Honolulu, HI, USA}
\acmBooktitle{Proceedings of the CHI Conference on Human Factors in Computing Systems (CHI '24), May 11--16, 2024, Honolulu, HI, USA}
\acmDOI{10.1145/3613904.3642816}
\acmISBN{979-8-4007-0330-0/24/05}

\usepackage{tabularx}
\usepackage{enumitem}
\usepackage{multirow}
\usepackage{graphicx}
\usepackage{subcaption}

\def \revision #1{\textcolor{black}{#1}}
\def \accept #1{\textcolor{black}{#1}}
\def \minor #1{\textcolor{black}{#1}}

\begin{document}

\title[Help Supporters]{Help Supporters: \revision{Exploring the Design Space of} Assistive Technologies \revision{to} Support Face-to-Face Help Between Blind and Sighted Strangers}

\author{Yuanyang Teng}
\affiliation{
    \institution{Columbia University}
    \city{New York}
    \state{NY}
    \country{USA}
}

\author{Connor Courtien}
\authornote{Author performed this work while at Columbia University.}
\affiliation{
    \institution{Hunter College}
    \city{New York}
    \state{NY}
    \country{USA}
}

\author{David Angel Rios}
\affiliation{
    \institution{Columbia University}
    \city{New York}
    \state{NY}
    \country{USA}
}

\author{Yves M. Tseng}
\affiliation{
    \institution{Columbia University}
    \city{New York}
    \state{NY}
    \country{USA}
}

\author{Jacqueline Gibson}
\affiliation{
    \institution{Columbia University}
    \city{New York}
    \state{NY}
    \country{USA}
}

\author{Maryam Aziz}
\authornotemark[1]
\affiliation{
    \institution{Duke University}
    \city{Durham}
    \state{NC}
    \country{United States}
}

\author{Avery Reyna}
\authornotemark[1]
\affiliation{
    \institution{University of Central Florida}
    \city{Orlando}
    \state{FL}
    \country{USA}
}

\author{Rajan Vaish}
\affiliation{
    \institution{Easel AI, Inc. }
    \city{Los Angeles}
    \state{CA}
    \country{USA}
}

\author{Brian A. Smith}
\affiliation{
    \institution{Columbia University}
    \city{New York}
    \state{NY}
    \country{USA}
}

\renewcommand{\shortauthors}{Yuanyang Teng, et al.}

\begin{abstract}
Blind and low-vision (BLV) people face many challenges when venturing into public environments, often wishing it were easier to get help from people nearby. 
Ironically, while many sighted individuals are willing to help, such interactions are infrequent. Asking for help is socially awkward for BLV people, and sighted people lack experience in helping BLV people. Through a mixed-ability research-through-design process, we explore four diverse approaches toward how assistive technology can serve as \textit{help supporters} that collaborate with both BLV and sighted parties throughout the help process. These approaches span two phases: the connection phase (finding someone to help) and the collaboration phase (facilitating help after finding someone). Our findings from a 20-participant mixed-ability study reveal how help supporters can best facilitate connection, which types of information they should present during both phases, and more. We discuss design implications for future approaches to support face-to-face help. \enlargethispage{10pt}
\end{abstract}




\maketitle

\section{Introduction}

Blind and low-vision (BLV) people face many challenges in their daily lives when venturing outside of their homes. These include understanding what is happening around them~\cite{easley_lets_2016, hurst_is_someone_there}, getting descriptions and directions of their environments~\cite{easley_lets_2016, williams_just_2014}, and deciphering what signs and other public displays around them say~\cite{thieme_i_2018}. 
BLV individuals often must struggle with these difficulties on their own and they would greatly appreciate it if getting help face-to-face from surrounding people could be easier.

As the popularity of remote volunteering platforms such as BeMyEyes \cite{be_my_eyes} shows, there are many people who are willing and eager to offer help to BLV people. The sighted volunteer base for BeMyEyes is thirteen times greater than the number of BLV users seeking help on the platform, according to their website in 2023. %(6,467,055 volunteers serving 481,569 BLV users)
%%% P: And yet, it doesn’t happen 
% 	P1 - BLV people face awkwardness to ask 
% 	P2 - Sighted people hesitate to offer 
% 	P3 - (social barrier) connection and (communication barrier) collaboration challenges 
% In short, it’s a social problem. We need to understand how to reduce social barriers.
Ironically, despite the plethora of opportunities for BLV people and surrounding sighted strangers to collaborate in public, the sad reality is that such collaborations seldom happen in times of need. We ask: \textit{Why does help for BLV people not happen more often?}
% Previous research revealed that it is due to the social challenges faced by both parties.  
% BLV people face social awkwardness reaching out for help because they cannot locate people accurately without vision and cannot gauge who is willing to help. Sighted people face social pressure when offering help because they hesitate when and how the BLV people would like to be helped. 
% BLV people face social awkwardness to ask for help as they cannot locate people without sight and gauge their willingness to assist. Sighted people face social pressure and are often hesitant to offer help due to uncertainty about the BLV person's needs and social etiquette preferences. In short, it’s a social problem. We need to understand how to reduce social barriers.
%When connecting, BLV people struggle to locate surrounding people for assistance, and sighted people are often unsure whether BLV people need help and do not want to guess incorrectly. After connecting, during collaboration, both parties face challenges in communicating effectively using non-visual descriptive language and navigating social etiquette preferences. In short, it is a social problem. Researchers need to understand how to reduce social barriers.

% BLV people often report being anxious and hesitant about the process of asking other people in their surroundings for help \cite{branham_is_2017, thieme_i_2018, i_want_to_figure}. Sighted people are often unsure when and how to offer help due to the misconception of BLV people's abilities \cite{guerreiro_airport_2019, thieme_i_2018}. 
Prior research suggests that social challenges are the root cause.
When it comes to finding help, which we call the \textit{connection phase}, BLV people often report being anxious and hesitant about asking other people in their surroundings for help \cite{hurst_is_someone_there, thieme_i_2018, i_want_to_figure}. Sighted people are often unsure when and how to offer help due to their misunderstanding of BLV people's needs and abilities \cite{guerreiro_airport_2019, thieme_i_2018}. 
When it comes to the process of helping itself, which we call the \textit{collaboration phase}, the two parties face additional challenges. Sighted people struggle to communicate effectively using non-visual descriptive language \cite{scheuerman_learning_2017, i_want_to_figure, lee_professionalpractice, kamikubo_support_2020}, leading to confusion for the BLV person.
% , such as  \yy{[Insert very short example here.]} 
Sighted people also struggle to honor the BLV person's boundaries and preferences~\cite{envision_how_nodate, missouri_how_nodate}, such as offering their elbow to guide them and not touching their white cane. \enlargethispage{10pt}

The problem of facilitating sighted help for BLV people is fundamentally a social problem rather than a technical one. In order to make headway on this problem, our community must understand how to reduce the social barriers for both parties.
%In short, it is a social problem. There presents an opportunity to solve this through better understanding how tech can be designed.

%%% S1: The recent interdependence framework reveals a possible solution (Ref DHH social paper language) - Opportunity for a new use case for Assistive technology for social xxx
%%% S2: AT might be a good solution/opportunity to help
Recent theoretical advances, specifically the interdependence framework ~\cite{bennett_interdependence_2018} and the concept of ``community-based accommodation'' ~\cite{kasnitz_politics_2020},  offer a promising direction for solving the social challenges faced by both BLV and sighted people. 
% The interdependence framework incorporates \textit{Assistive Technology} (AT) in the same environment together with BLV people and other surrounding people, arguing for equal competence and agency from all parties. 
% "Community-based accommodation" suggests that access to assistive accommodations can equally benefit both BLV individuals and sighted individuals. There is an opportunity to create a new use case that leverages AT in a social context and bridges social barriers between these two groups.
% S1. Community-based accommodation, which argues that AT should be thought of as a tool that can help both parties---not only people with disabilities---and that both parties should have the ability to request help that they need from AT.
% S2. Interdependence: AT is a connecting partner in a social environment
% ...the interdependence framework, which argues that the goal of making the world accessible to people with disabilities is not necessarily a job for assistive technology specifically but can rather be viewed as a shared goal that people with disabilities, assistive technology, surrounding people, and environmental infrastructure can collaborate towards achieving.  
% Assistive Technology Act of 1998 defined Assistive Technology (AT) as tools that are used to assist individuals with disabilities \cite{act_1998}. 
The concept of community-based accommodation~\cite{kasnitz_politics_2020} expands the legal definition of assistive technology ~\cite{act_1998}, arguing that assistive technology should not only provide people with disabilities with support, but also people without disabilities so that the latter can better work with the former.
% as tools that are used to assist individuals with disabilities by arguing that AT should also assist people \textit{without disabilities} to further improve people with disabilities' experiences \cite{kasnitz_politics_2020}. 
% It argues that not only BLV people can request help from AT, but also sighted people who need accessibility assistance in order to interact with BLV people should also be supported by AT.
Furthermore, the interdependence framework~\cite{bennett_interdependence_2018} argues that the goal of making the world accessible is not solely shouldered by assistive technology but can be a shared goal toward which people with disabilities, technology, surrounding people, and environmental infrastructure collaborate to achieve. Recent social computing research~\cite{winkler_alexa_2019, elshan_lets_2020, neto_robot_classrooms} has explored leveraging technologies as social agents to improve teaching and other team collaboration, but to the best of our knowledge, no work has explored how assistive technology can support face-to-face help between BLV and sighted strangers.

% \revision{We took the opportunity to articulate community-based accommodation by exploring four diverse approaches toward how assistive technology can serve as \textit{help supporters} that collaborate with both BLV and sighted parties throughout the help process.}

\revision{In this work, we undergo a research-through-design process ~\cite{gaver_researchthroughdesign, zimmerman_researchthroughdesign} to imagine and explore the design space for} a new category of assistive technologies that we call \textit{\textbf{Help Supporters}}, whose purpose is to accompany both BLV and sighted people in person to address the social barriers \accept{that prevent} help.
We frame the design goals in terms of the two phases of help: the \textbf{\textit{connection phase}} where the BLV person and sighted stranger establish contact, and the \textbf{\textit{collaboration phase}} where the two parties work together in person.
% We call the initial phase of establishing contact between a BLV person and a sighted stranger the \textbf{\textit{Connection Phase}}, and we call the subsequent process of the two parties working together the \textbf{\textit{Collaboration Phase}}. 
During the connection phase, the two parties must identify each other in the environment, greet each other, and determine whether the sighted individual is capable and available to meet the needs of the BLV person.
% After that, both parties work together to fulfill a help request, we call this the \textbf{\textit{Collaboration Phase}}. 
During the collaboration phase, the parties must communicate effectively and respect each other's boundaries. Additionally, the sighted helper must be aware of the BLV individual's social etiquette preferences.

%%% S1: We develop and systematically investigate several approaches for facilitating help, focusing on several important questions:
%%% 	RQ1: Who should initiate the connection?
%%% 	RQ2: Where should messages be displayed?
%%% 	RQ3: Who should be aware of and have control over the messages?
%%% 	RQ4: What 
%%% 	RQ5: How much agency

\revision{Our team of mixed-ability co-authors conducted an internal co-design process to identify six major design attributes for \textit{help supporters} (Table~\ref{fig:design_axes}), each of which could be assigned different settings to result in very different \textit{help supporter} designs. Through an iterative design process, we arrived at four prototypes (two for each phase) to explore different regions of the design space. These prototypes are unconventional but allow us to explore the following major research questions: } \enlargethispage{15pt}
%To uncover insights about the design space of help supporters, we develop four prototypes as technology probes ~\cite{Hutchinson2003} (two for each phase of help), which represent different approaches for supporting help. We explore the following research questions:

%\vspace{0.5em}  
% \textbf{How should technology help during the Connection Phase? }
\noindent\textbf{Connection phase:}
%\vspace{0.3em}

% \begin{tabular}{{p{0.05\textwidth}p{0.85\textwidth}}}
%  \textbf{RQ1.} & Who, BLV people or sighted strangers, should initiate the connection?  \\
%  \textbf{RQ2.} & How may \textit{Be There Help Agents} support both BLV and sighted people to initiate the connection?\\
% \end{tabular}

% 
% [RQ1:] How should the AT help during the connection phase?
% Should the AT agent encourage in-person requests, or should it foster an app-based platform for requests?   <--- We’re going to reveal users’ thoughts after trying these possibilities, then come to some conclusions.
% What type of information should the AT facilitate to each party during the connection phase?
\noindent 
\begin{tabular}{{p{0.08\linewidth}p{0.8\linewidth}}}
\textbf{RQ1.} & Should \textit{help supporters} encourage BLV people to make face-to-face requests or app-based requests for help from nearby strangers?  \\
\textbf{RQ2.} & What types of information should help supporters give BLV and sighted people about each other for the purpose of connecting?\\
\end{tabular}

%\vspace{0.4em}  
% \textbf{How should technology help during the Collaboration Phase?}
%\newpage
\noindent\textbf{Collaboration phase:}
\vspace{0.3em}

\noindent 
\begin{tabular}{{p{0.08\linewidth}p{0.8\linewidth}}}
 \textbf{RQ3.} & What types of information should help supporters provide to sighted helpers during help?  \\
 \textbf{RQ4.} & Where should help supporters situate the information during help?  \\
 % \textbf{RQ5.} & What abilities should \textit{Be There Help Agents} have? \\
\end{tabular}  
\vspace{0.5em}  

% Our design probes for answering these research questions
% To answer these research questions, we evaluate our four 
% technology probes 
% approaches
% (two approaches for each phase) to evaluate 
% via a user study with 20 participants (10 mixed pairs of sighted and BLV participants). 
\newpage
Our objective with the two connection phase \revision{prototypes} is to explore how technology should help during this phase (RQ1) and to uncover the information that should be exchanged between BLV individuals and nearby sighted strangers (RQ2). 
% To investigate \textbf{RQ1} and \textbf{RQ2}, we implemented two approaches.
% \textit{Person-Finder Glasses} takes a BLV initiative approach, consisting of a HoloLens app that can detect human figures to help BLV people find and approach sighted strangers.
Our first \accept{prototype}, the \textbf{\textit{Person-Finder Glasses}}, encourages face-to-face requests for help. It uses computer vision and audio cues to enable BLV people to detect and approach others around them. 
% \textit{Volunteer Platform} takes a sighted initiative approach, consisting of a mobile app that allows sighted strangers to accept help requests from BLV people so that they can reach out to offer help.
Our second \accept{prototype}, the \textbf{\textit{Volunteer Platform}}, explores \accept{the use of} a mobile app platform to facilitate requests and offers for help. The two parties are matched via the app before they meet in person. 

%For the collaboration phase, the objective is to explore how AT can help sighted people match BLV people's preference for nonvisual descriptive language and social etiquette.
% For the \textit{Collaboration Phase}, the objective 
Our objective with the two collaboration phase \revision{prototypes} is to explore how technology can \accept{support} BLV and sighted people to communicate better during the help process, once they are already together in person (RQ3 and RQ4). 
%To investigate \textbf{RQ3} and \textbf{RQ4}, we designed two approaches to explore the placement of the corrective messages, and the agency of the technology and the users:
% To investigate \textbf{RQ3} and \textbf{RQ4}, we designed
The two \accept{prototypes} explore the types of information that should be facilitated (RQ3) and how that information should be placed for the sighted helper (RQ4).
%\textit{Prototype C} takes a public display and low agency approach. It consists of a wearable display screen that broadcasts the corrective messages in a fun friendly manner visible to all surrounding strangers. Its low agency would only suggest corrections, and it is up to the users to ignore or follow the messages. 
The \textbf{\textit{Pictorial Display}} \accept{prototype} takes a public display approach with an image-forward format. It consists of a wearable screen that broadcasts the messages in a lighthearted pictorial format visible to all surrounding strangers. 
%\textit{Prototype D} takes a private display and high agency approach. It consists of a speech-to-text mobile app on the sighted helper’s phone that commands them to improve their descriptive language. Its high agency design shows yellow highlights on the erroneous parts of the language until users make corrections. 
The \textbf{\textit{Vague Directions Flagger}} \accept{prototype} takes a private display approach with a more specific form of text feedback. It is a smartphone app that runs on the sighted helper’s smartphone, identifying vague directions and descriptions that the sighted person gives (by transcribing their speech) and prompting the sighted person with ways to improve them.

Through a user study with 20 participants (10 mixed pairs of sighted and BLV participants), we \revision{uncovered insights about several key design aspects of} help supporters~\revision{ (RQ1–RQ4), such as face-to-face requests vs.\ app-based requests for help (RQ1) and the types of information that help supporters should offer users during help (RQ3)}.
%, summarized in Table~\ref{fig:table_findings}. 
We found that \accept{during the connection phase,} having app-based help requests reduces social pressure for both BLV and sighted people, making both more willing to join together. BLV people prefer to know their helpers' level of knowledge and time availability before asking for help. During the collaboration, sighted helpers seek real-time feedback on their performance and encourage help supporters to educate them when needed on the spot, yet hide their mistakes from other sighted people. We conclude our work with design implications for future efforts in technology-mediated \accept{mixed-ability} help. \enlargethispage{10pt}

%%%% 1_v2_introduction.tex ends here %%%%

%%%% 2_v4_related_work.tex starts here %%%%

\section{Related Work}

\subsection{Collaboration Between BLV and Sighted Individuals}
\label{sec:related_collab}

%1. BLV and Sighted collaboration is important and desired. Reasons - learning from each other, effective help, community inclusion
%2. social challenges, and how current solutions are not effective. 

Researchers have investigated how BLV and sighted individuals collaborate for everyday activities across various scenarios such as navigation \cite{williams_just_2014, i_want_to_figure, williams_pray_2013}, 
%personal safety in public \cite{branham_is_2017}, 
shopping \cite{yuan_i_2017}, workplaces \cite{branham_invisible_2015, metatla_cross-modal_2012, naraine2011social}, classrooms \cite{neto_robot_classrooms}, and households \cite{branham_collaborative_2015}. 
The value of their collaboration extends beyond enabling BLV people to achieve independent living and encompasses fostering their active social integration with society as equal individuals ~\cite{yuan_i_2017}, improving cohabitation in shared space ~\cite{branham_collaborative_2015, branham_invisible_2015}, and promoting increased mutual \accept{understanding} ~\cite{flatla2012so}. 
%Prior research revealed that collaboration between sighted and BLV individuals is not easy and have three areas of challenges: social barrier, misperceptions of abilities and needs, behavioral and semantic misunderstandings. 
Previous research has revealed that collaboration between sighted and BLV individuals is far from straightforward and is characterized by challenges across three areas: social barriers, misperceptions of abilities and needs, and behavioral and semantic misunderstandings.
% The current state of research has attempted to address these challenges by providing BLV people with visual information, providing sighted people with guidelines, and having assistive technologies replace human help. 
% However, little research has investigated approaches to solve these challenges during sighted and BLV collaboration in situ.
% \yy{(add precursor to gaps) methods such as O\&M training, facial recognition, and advocacy training have been proposed and experimented with, but the challenges remain stubborn. GAP is that none of the methods solve the in-situ real-time problem.}

%Social barrier faced by BLV people stemmed significant hesitation for connecting with others for assistance. 
%Social barriers faced by BLV people stem from significant hesitation in connecting with others for assistance.
Regarding social barriers, BLV individuals express concerns about inconveniencing sighted helpers and subjecting them to social pressure~\cite{i_want_to_figure}, and they often hesitate and feel awkward reaching out to strangers for help. 
% Prior research also revealed that 
BLV individuals also often lack awareness of other people in their surroundings and are anxious about identifying potential helpers when they need help~\cite{thieme_i_2018, hurst_is_someone_there}. 
% In addition, BLV individuals express concerns about inconveniencing sighted helpers and subjecting them to social pressure~\cite{i_want_to_figure}.
%O&M training????
To address the difficulty of locating and identifying other people in the environment, several works have explored using computer vision on wearable camera feeds to give BLV people information about nearby people \cite{lee_pedestrian_2020, morrison_peoplelens_2021, Grayson_adynamicai, jin_smart_2015, stearns_automated_2018}. Most of these works, however, focus on enabling BLV people to recognize people they already know rather than to meet strangers. 
%Only Lee et al. \cite{lee_pedestrian_2020} reported findings from both BLV and sighted users' perspectives, but their research focused on privacy concerns of using wearable cameras. 
They also do not investigate \accept{or compare with }other means of connecting the two parties (such as via an app-based platform).

Regarding the misperceptions of BLV individuals' abilities and needs,
%of BLV individuals remains a common challenge faced by sighted individuals \cite{thieme_i_2018}.
sighted strangers often mistakenly categorize BLV individuals under a uniform disability label, leading to unintentional behaviors that do not align with the actual abilities of BLV individuals~\cite{thieme_i_2018}. This can manifest as actions such as speaking louder, slowing down, or offering unnecessary assistance like transportation when walking is feasible~\cite{guerreiro_airport_2019, thieme_i_2018}.
To address this challenge, BLV-serving organizations have incorporated guidelines on their websites to educate sighted individuals. These guidelines recommend sighted individuals speak normally, talk to BLV people directly~\cite{missouri_how_nodate, envision_how_nodate,wisconsin_health_2010}, and ask if assistance is needed~\cite{wisconsin_health_2010, blind_perkins_1970}. 
Despite this, during everyday encounters, many sighted people do not know how to interact with BLV individuals and help them effectively. We hypothesize that this is due to the lack of prior exposure to these guidelines. There is a lack of an in-situ and real-time approach to support sighted people in their interaction with BLV people---a goal of help supporters during the collaboration phase.
Lastly, regarding behavioral and semantic misunderstandings,
BLV people prefer verbal descriptions from others that align with their non-visual perception of the environment~\cite{i_want_to_figure}, yet sighted people often describe things in visual ways (e.g., pointing ``Over there!'', ``It's just past the blue door.'').
% Existing research has highlighted that BLV individuals express a preference for verbal descriptions from others that align with their non-visual perception of the environment ~\cite{i_want_to_figure}. In a field observation, Williams et al. ~\cite{williams_just_2014} pinpointed challenges encountered by both BLV and sighted individuals when communicating navigation assistance. 
Other communication challenges include confusing phrases, omitted information, and vague orientation descriptions~\cite{williams_just_2014}, prompting researchers to develop guidelines for giving directions to BLV people
% In a similar vein, specific research has focused on examining the language employed to convey directions to BLV individuals 
~\cite{scheuerman_learning_2017, perez_assessment_2017, nicolau_blobby_2009, lee_professionalpractice, kamikubo_support_2020}. 
Furthermore, there exists a tendency for sighted individuals to frequently misinterpret BLV people's behavior, often mistaking the act of following an edge with a white cane (to maintain orientation) as running into obstacles~\cite{williams_just_2014, i_want_to_figure}.
Our two collaboration-phase prototypes explore how \revision{assistive technology} can address these types of misunderstandings\revision{, and users' attitudes toward assistive technology working in this way}.

\subsection{\revision{BLV People’s Current Practices for Navigating and Sensemaking}}
\label{sec:related_currentpractices}

BLV people navigate and make sense of their surroundings through a set of unique practices that often involve their non-visual senses, orientation and mobility skills, and cues and guidance from other people ~\cite{thieme_i_2018}. 
It is important to note that, while BLV people’s process for navigating and sensemaking is different from sighted people’s process, it is equally valid and just as effective.

Thieme et al. ~\cite{thieme_i_2018} describe a representative scenario of a BLV person navigating through an airport that highlights several methods that BLV people employ to navigate successfully. The BLV person first uses a magnifier to read signage. While trying to find the right signage, they spot a person wearing bright yellow clothes, whom they think is a security staff member that they can approach with a question. The BLV person unintentionally frames the question in a confusing way due to their vision, but is still able to successfully gather useful information from the brief conversation. Following that, the BLV person learns more about the airport's layout by wandering around and reading more signs. Eventually, they find the signage pointing to their departure gate.
%Trying to find their departure gate in an airport, the BLV person tried to locate and read signage using a magnifier. After getting close to the gate number but could not find the exact gate, the BLV person located a person in bright yellow color whom he interpreted as a staff. Although the way 

\revision{The authors identify several opportunities for assistive technology to support BLV people’s current practices: by enabling BLV people to better identify people around them and choose who to interact with, by fostering a shared understanding of other people’s actions, and by considering existing social relationships between them.
Through this research-through-design process, our objective is to explore and compare different approaches that represent these opportunities for assistive technology to learn more about what designs work best and what user attitudes and preferences towards such technologies are. 
}

%\subsection{Technologies as Assistive and Collaborative Agents}
\subsection{Assistive Technologies Used by BLV and Sighted Individuals}
\label{sec:related_usedby}
%\outline{TODO: on theoretical framework -  definition of assistive technology, interdependence, and community tech theory. and a sentence leading to prior work examples and gaps.}

The Assistive Technology Act of 1998 \cite{act_1998} originally defined assistive technology as tools used to aid individuals with disabilities. 
Since its enactment, this definition has evolved and expanded due to both theoretical research advancement and assistive technology design progress to include considerations for individuals without disabilities. This evolution marks a significant step forward in human rights that rejected disabilities as a medical condition and embraced a social model of disability \cite{oliver1990individual}. 
% from viewing disability as social challenges rather than medical conditions
% \yy{add why extending to include people without disability is important. medical -> social shift}

% Multiple thories in research have expanded definition of assisitve technology to include people without disabiliites in different ways.
% Community-based accommodation framework argues that AT should also assist people without disabilities to gain access to people with disabilities \cite{kasnitz_politics_2020}. Social accessibility \cite{shinohara_socialaccessibility} explored the use of assisitive technology in the presence of others sheding light to the issue of social acceptance of assistive technology. Interdependence framework \cite{bennett_interdependence_2018} further established the view that assistive technology, people with disabilities, other surrounding people, and the environment are in a partnership together to solve accessiblity problems and making the world accessible to all. 

Theoretical frameworks in research have broadened the scope of assistive technology to include individuals without disabilities in multiple facets. 
The community-based accommodation framework argues that assistive technology should facilitate access not only for individuals with disabilities, but also for those without disabilities to interact with individuals with disabilities \cite{kasnitz_politics_2020}. Social accessibility studies the use of assistive technology in the presence of others, shedding light on the issue of social acceptance of such technology \cite{shinohara_socialaccessibility}. 
The interdependence framework further solidifies the view that assistive technology, individuals with disabilities, other people in proximity, and the physical environment form a partnership to address accessibility challenges and make the world universally accessible \cite{bennett_interdependence_2018, vincenzi_interdependence}.
Our research acts upon the existing theories to design assistive technologies for both BLV and sighted users, in a context involving all the stakeholders: sighted helpers, BLV helpees, the \textit{help supporter} prototypes, and the physical environment.

% Building upon these theoretical foundations, the design of assistive technology has incorporated individuals without disabilities as active users.

% In scenarios where assistive technology is used in the presence of others, its design considers its impact on social interactions. This is particularly relevant for individuals with disabilities using assistive technology in shared spaces and activities \cite{branham_collaborative_2015, branham_invisible_2015}.
% Assistive technology has also been designed for collaborative usage by individuals with mixed abilities. Examples include interdependence guidance in physical navigation\cite{vincenzi_interdependence}, the integration of robots in classrooms with students of differing abilities \cite{neto_robot_classrooms}, and the development of multimodal user interfaces that facilitate collaboration between BLV and sighted individuals \cite{plimmer2008multimodal, savidis1995developing}.

% \yy{
% WHY: social accessiblity, medical accessible, 
% GAP: no work has dealt with help between BLV and Sighted }

In the domain of remote sighted assistance, tools like AIRA~\cite{aira}, VizWiz~\cite{bigham_vizwiz_2010}, and BeMyEyes~\cite{be_my_eyes} leverage crowdsourcing to connect BLV individuals with sighted volunteers or professionals for assistance. 
However, sighted individuals providing assistance through these systems encounter the challenge of not sharing a common visual perspective with the BLV individuals they are aiding.
To address this gap, 
%researchers have introduced tools to aid sighted individuals in providing more accurate navigation guidance to BLV users, such as creating remote maps of BLV user's environment for sighted helpers to better understand BLV's environment \cite{helping_helpers}. 
researchers have introduced tools aimed at aiding sighted individuals in delivering more accurate navigation guidance to BLV users. This includes creating remote maps of the BLV user's environment, which enables sighted helpers to gain a better understanding of the physical surroundings and nature of tasks faced by BLV individuals~\cite{helping_helpers}.
Despite this, when compared to in-person co-located assistance, these remote assistance tools have limitations. They do not facilitate the physical co-presence of sighted individuals and the environment alongside BLV individuals. This absence hinders an interdependent social environment that fosters collaborative interaction and inclusivity. 
Our present research explores how \revision{assistive} technology \revision{might} facilitate face-to-face help and mutual understanding\revision{, as well as users' attitudes toward assistive technology working in this way.}

\begin{table*}
    \centering
    
    \caption{
    \revision{Summary of the six design attributes identified during our internal co-design process. We group these by the two phases of help (the Connection Phase and Collaboration Phase).}
    }
    \includegraphics[width=0.75\textwidth]{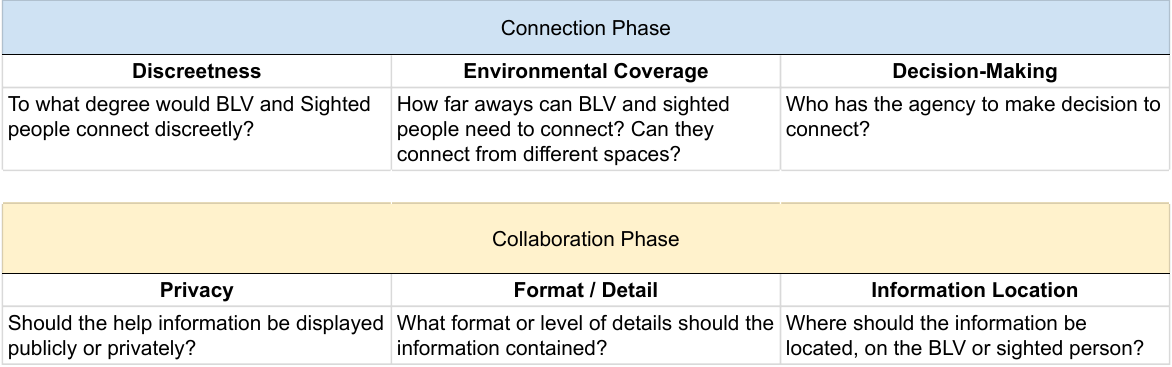}
    \label{fig:design_axes}
    \Description{Summary of the six design attributes identified during our internal co-design process. The table contains three attributes for the Connection Phase: discreetness, environmental coverage, and decision-making. It also contains three attributes for the Collaboration Phase: privacy, format / detail, and information location. Below each of the six attributes is a more thorough description of that attribute. For Discreetness, for example, the description is “To what degree would BLV and Sighted people connect discreetly?” Each of these cells is rendered as plain text.}
\end{table*}

\begin{table*}

    \caption{
    \revision{Summary of how the four prototypes map to the design attributes. We group these by the two phases of help (the Connection Phase and Collaboration Phase).}
    }
    \includegraphics[width=0.9\textwidth]{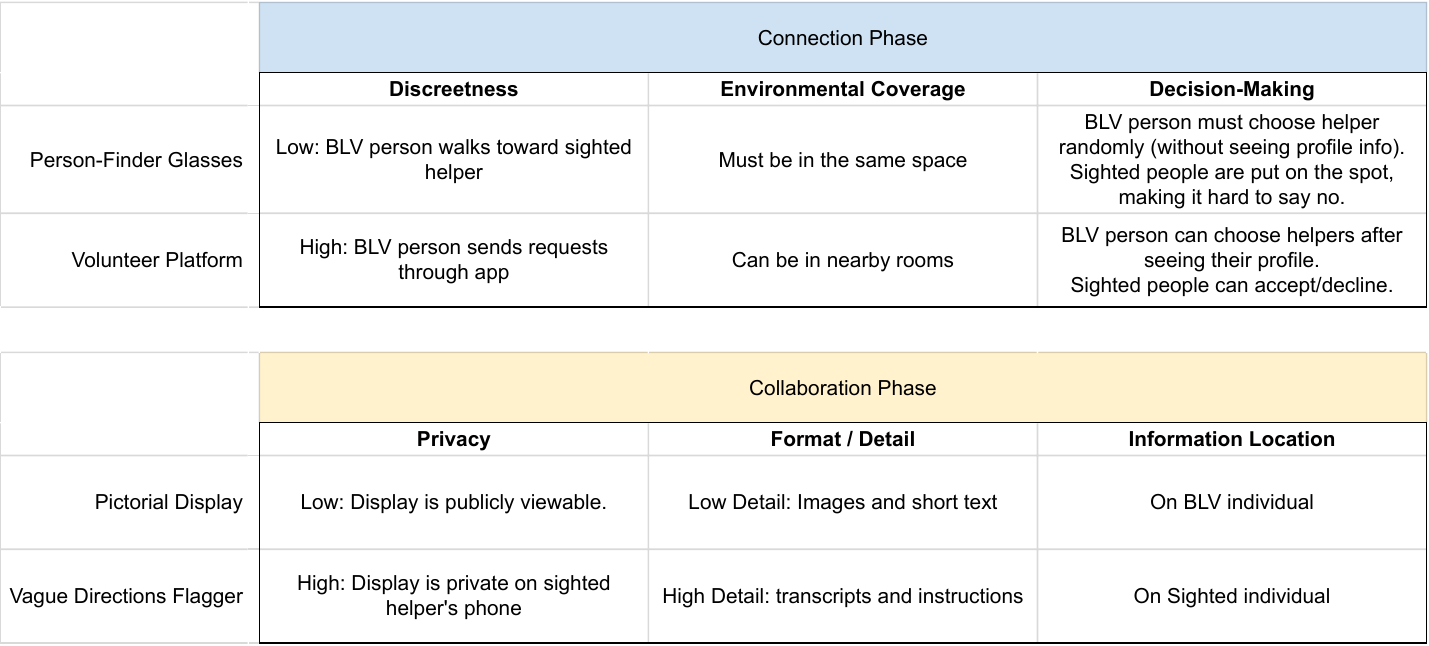}
    \label{fig:design_axes_withvalues}
    \Description{Summary of how the four prototypes map to the design attributes. There are two sub-tables stacked one above the other. The first sub-table is for the Connection Phase and has two rows, one for Person-Finder Glasses and one for Volunteer Platform. It has three columns labeled “Discreetness,” “Environmental coverage,” and “Decision-making.” The cells of the table describe the attribute values for each of the two Connection Phase prototypes. The second sub-table is for the Collaboration Phase prototypes and has two rows as well: one for Pictorial Display and one for Vague Directions Flagger. It has three columns labeled “Privacy,” “Format / Detail,” and “Information Location.” The cells of the table describe the attribute values for each of the two Collaboration Phase prototypes.}
\end{table*}

\begin{table*}
    \centering
    
    \caption{
    Summary of the \textit{Help Supporter} prototypes for the Connection Phase and Collaboration Phase, 
    and how each of the four stakeholders\revision{---the \textit{Help Supporter}, the sighted individual, the BLV individual, and the environment---}interact to \revision{support the collaborative process of help via the interdependence framework~\cite{bennett_interdependence_2018, vincenzi_interdependence}}.
    }
    \includegraphics[width=1\textwidth]{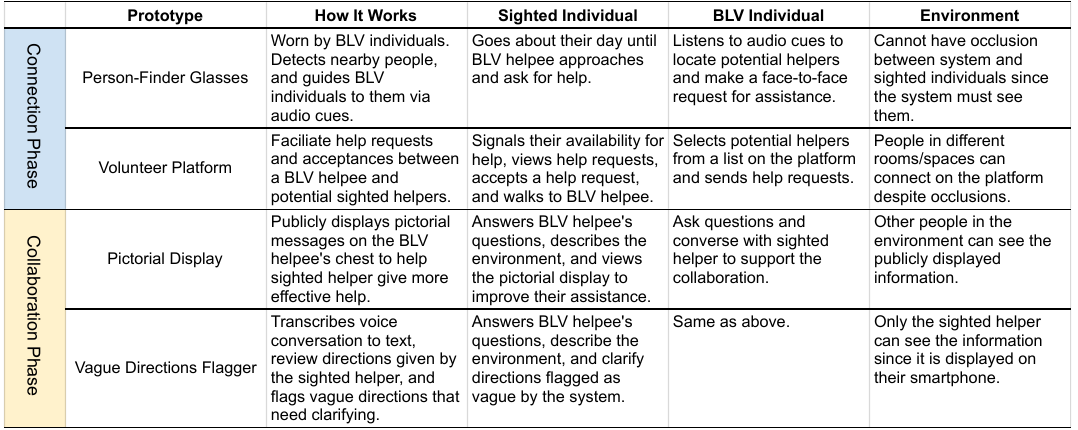}
    \label{fig:summary_prototypes}
    \Description{Summary of the Help Supporter prototypes for the Connection Phase and Collaboration Phase. The table contains four rows, one for each of the four Help Supporter prototypes. The first two rows correspond to the prototypes for the Connection Phase, and the last two rows correspond to the prototypes for the Collaboration Phase. The table also contains five columns, the first column is the prototype name, the second column describes How it Works, and the third to the fifth columns correspond to each of the three relevant stakeholders: the sighted individual, the BLV individual, and the environment. The cells of the table describe how each of the four prototypes interacts to bring BLV and sighted strangers together for help and facilitate communication improvements during a help session.}
\end{table*}

\begin{figure*}
    \centering
    \vspace{5mm}
    \includegraphics[width=0.8\textwidth]{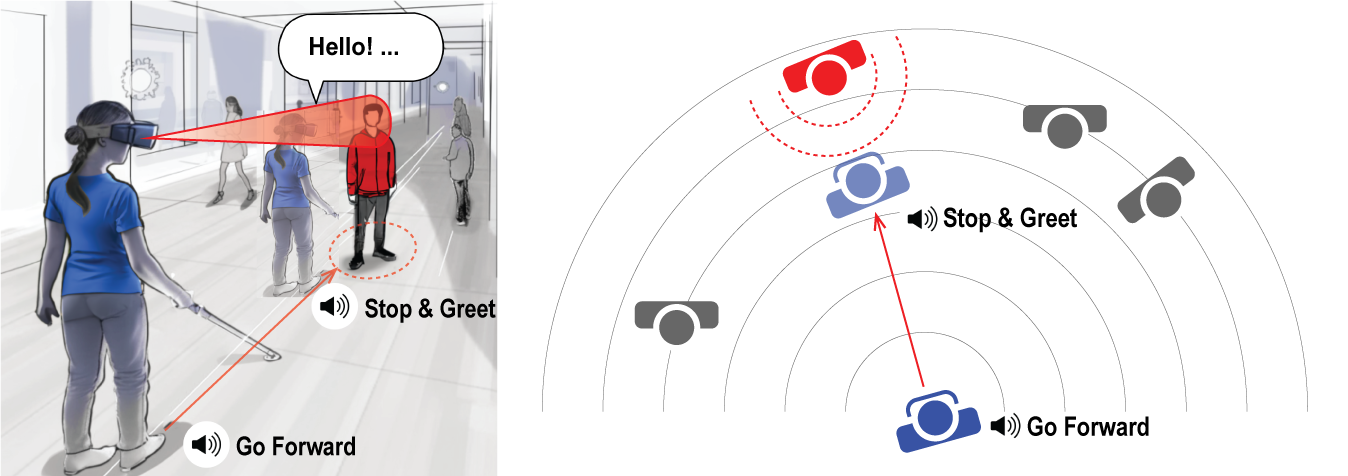}
    \caption{\textbf{\textit{Person-Finder Glasses}}. Left: A BLV user (in blue) wearing the \textit{Person-Finder Glasses} to locate and greet a nearby stranger (in red) in a public environment. Right: A plan view diagram showing how the \textit{Person-Finder Glasses} support the BLV user (in blue) to detect a nearby stranger (in red) and display auditory cues.}
    \label{fig:system_personfinder}
    \Description{Two diagrams depicting how the Person-Finder Glasses prototype works. On the left, a BLV person in blue is wearing the Person-Finder Glasses on their head in a public environment, and a sighted person in red is located a few steps in front of them. They are using the auditory feedback that the Glasses give them to determine when to go forward towards the sighted person, and at what point they are close enough to that sighted person to stop moving forwards. On the right, a bird’s-eye-view diagram depicts how the Person-Finder Glasses enables a BLV individual to locate a person in their would-be field-of-vision. Concentric gray arcs are depicted emanating from the BLV individual’s Person-Finder Glasses, and extend towards the sighted person that they are approaching. These arcs represent the person-detection capabilities of the Glasses.}
\end{figure*}

\begin{figure}
    \centering
    \includegraphics[width=0.47\textwidth]{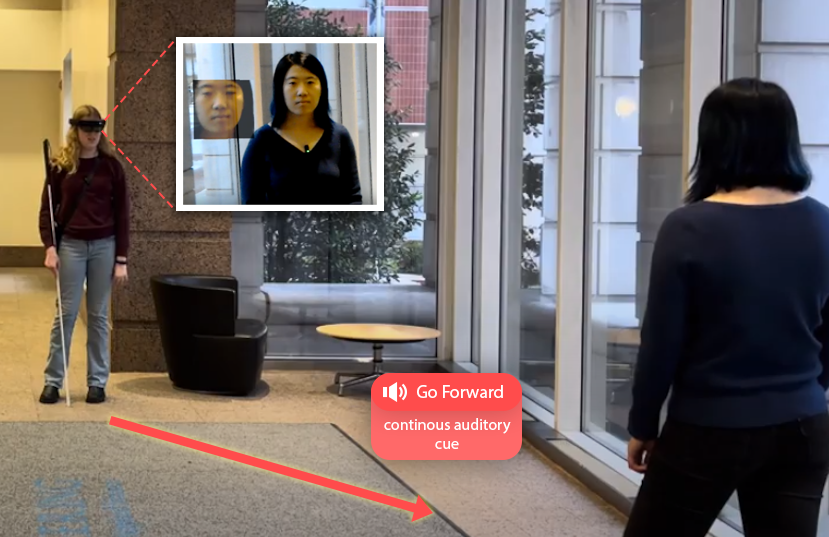}
    \caption{\revision{\textbf{\textit{Person-Finder Glasses implementation}.} \newline A blind user looks around with the Person-Finder Glasses to find help. When the system detects a face \textit{(inset, with an overlay indicating a detection)}, it plays a continuous spatial audio cue to guide the wearer to the other person.}}
    % Left: Screenshot of the BLV user's perspective while using the \textit{Person-Finder Glasses}. Right: The BLV user is receiving a continuous auditory cue to guide them to the sighted user. As long as they continues to face the sighted user as they approach, they will continue to hear an auditory cue.  }}
    \label{implement_1}
    \Description{Person-Finder Glasses implementation. A photograph depicting a blind person with a white cane who is wearing a head-mounted display and looking toward a sighted person across a large lobby. An inset image shows the first-person view from the head-mounted display, depicting the sighted person’s face centered in the frame. A red arrow from the blind person to the sighted person is overlaid on the diagram with an icon indicating a continuous auditory cue, suggesting that the display is playing a sound from the spatial direction of the sighted person.}
\end{figure}

\begin{figure*}
    \centering
    \includegraphics[width=0.9\textwidth]{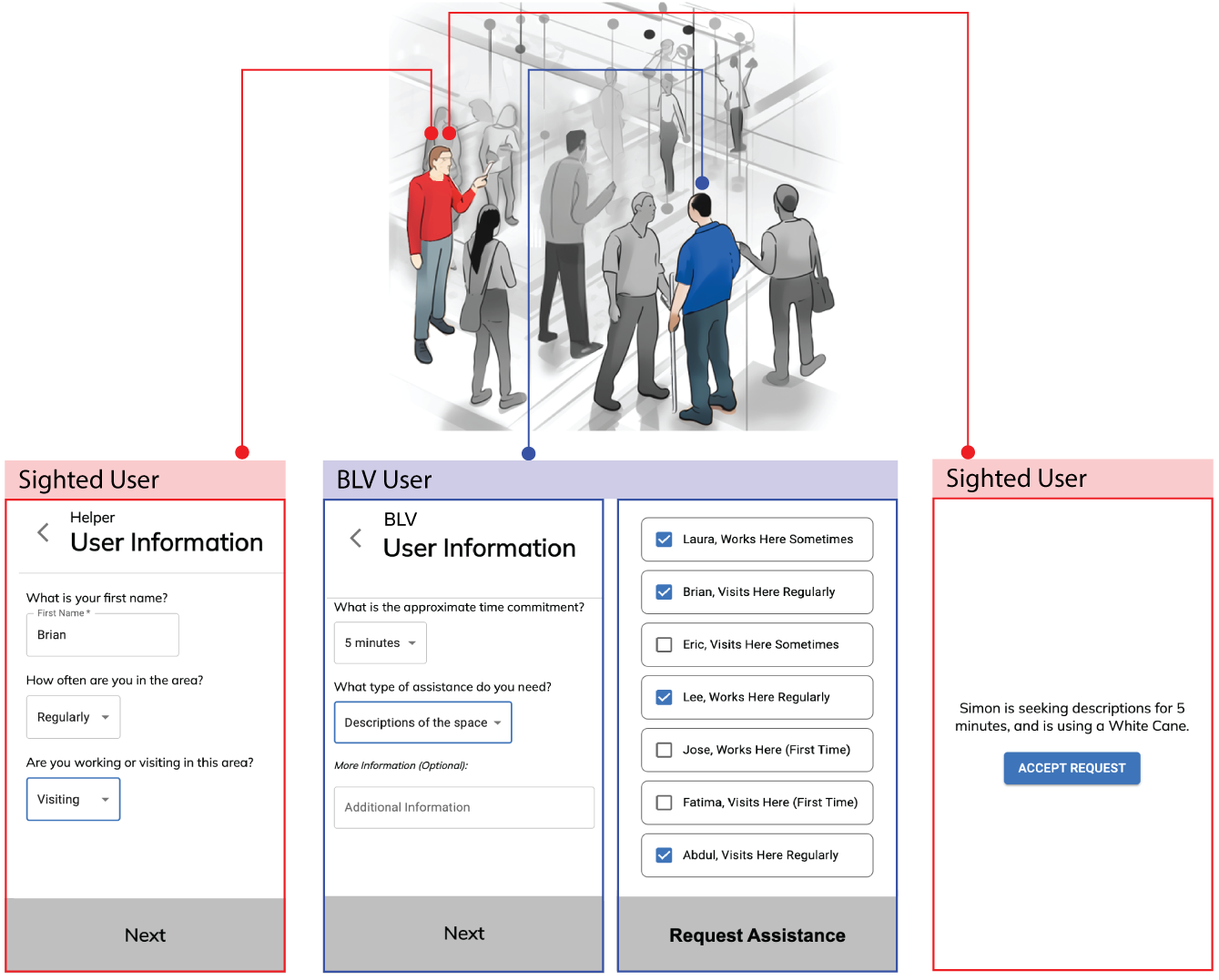}
    \caption{\textbf{\textit{Volunteer Platform}}. Top: A BLV user (in blue) using the \textit{Volunteer Platform} to connect with one of the nearby helpers (in red). Bottom: Screenshots of the \textit{Volunteer Platform} mobile app. The sighted user (in red) provides basic information about themselves, and the BLV user (in blue) sends help requests to a list of nearby helpers.}
    \label{fig:system_volunteerplatform}
    \Description{A diagram depicting how the Volunteer Platform prototype works, accompanied by four screenshots of the Volunteer Platform app interface. The diagram depicts a crowded public environment, with two people highlighted: a BLV person in blue, and a sighted helper in red. In the screenshots of the app interface, the first screenshot shows a form for the sighted volunteer to input their basic information (first name, how often they are in the area, whether they are working or visiting the area), the second screenshot shows a form for the BLV user to input their basic information(approximated time needed for help, type of help needed, optional more information), the third screenshot shows a list of nearby helpers for the BLV user to choose from, the fourth screenshot shows a help request on the sighted helper’s phone with an accept request button.}
\end{figure*}

\begin{figure}
    \centering
    
    \begin{subfigure}{1\linewidth}
    \includegraphics[width=1\textwidth]{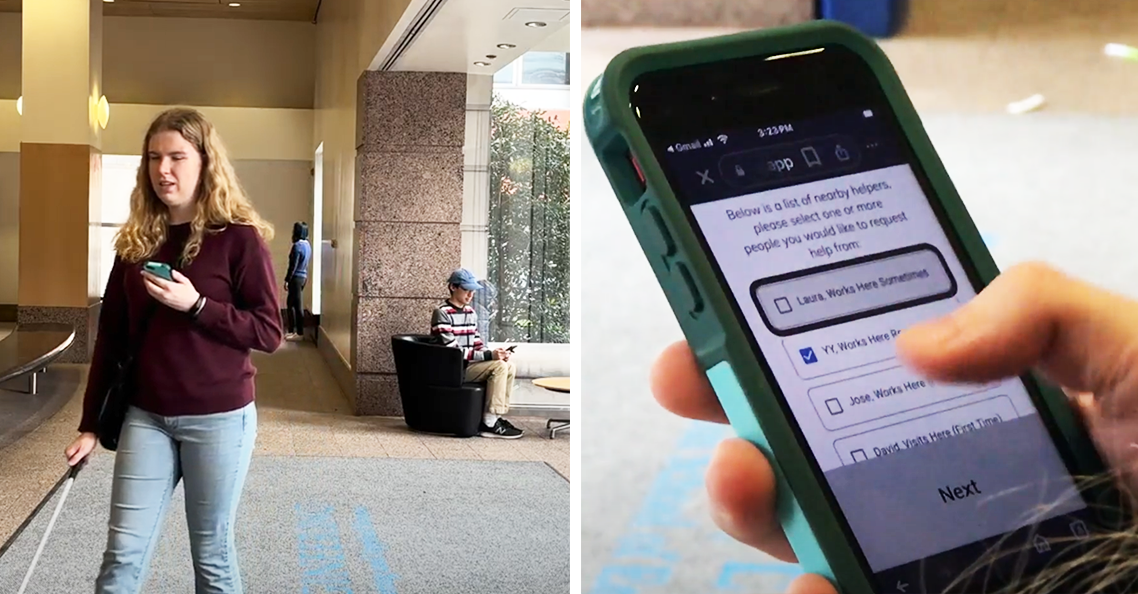}
    \caption{Some sighted people are nearby in the background \textit{(left)} as the BLV user chooses possible helpers on the \textit{Volunteer Platform} \textit{(right)}.}
    \end{subfigure}

    \hfill

    \begin{subfigure}{1\linewidth}
    \includegraphics[width=1\textwidth]{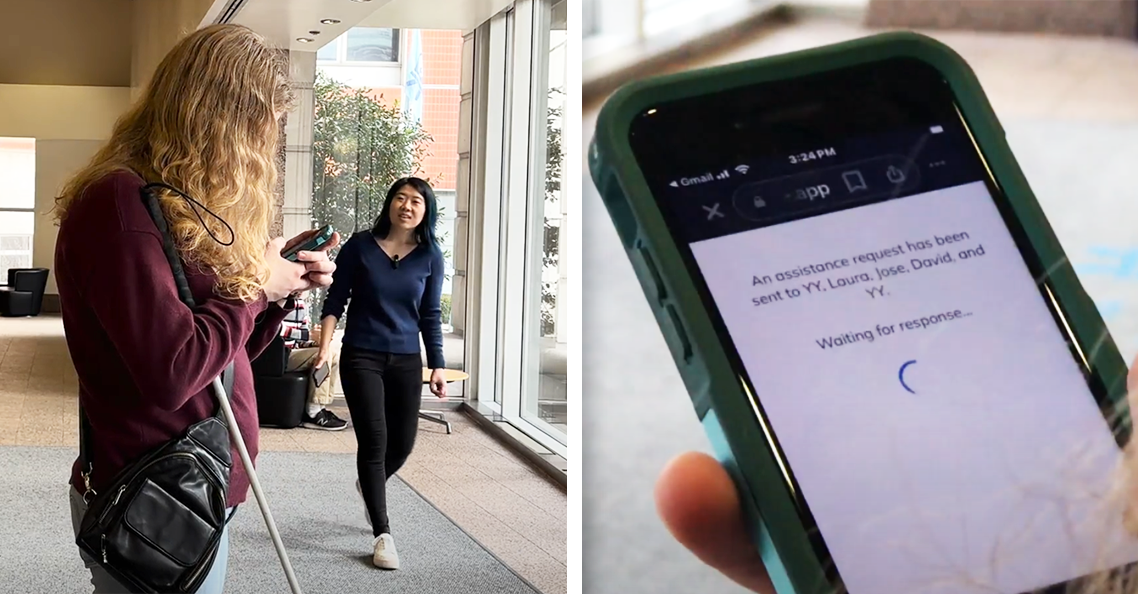}
    \caption{A sighted helper has accepted the request and is now approaching the BLV user \textit{(left)}, and the BLV user is notified via the app that help request has reached the volunteer helpers \textit{(right)}.}
    \end{subfigure}

    \caption{\revision{\textbf{\textit{Volunteer Platform implementation.}} 
    % (A) The matching process begins. A potential sighted helper waits in the background \textit{(left)} as the BLV user chooses possible helpers on the \textit{Volunteer Platform} \textit{(right)}. (B) The sighted helper has accepted the request and is now approaching the BLV user. 
    }}
    \label{implement_2}
    \Description{Two sets of photographs (a) and (b). In photograph (a), at left, a blind person holding a white cane is shown using their smartphone with some sighted people nearby in the background; at right, a close-up view of their smartphone’s display is shown, revealing a list of volunteer helpers’ first names and profiles with checkboxes next to each one. In photograph (b), at left, a sighted person begins to approach the blind person;  at right, a message reading “Assistance request has been sent to (selected volunteer helpers’ names) ” is seen in a close-up view of the blind person’s smartphone screen.
}
\end{figure}

\begin{figure*}
    \centering
    \includegraphics[width=1\textwidth]{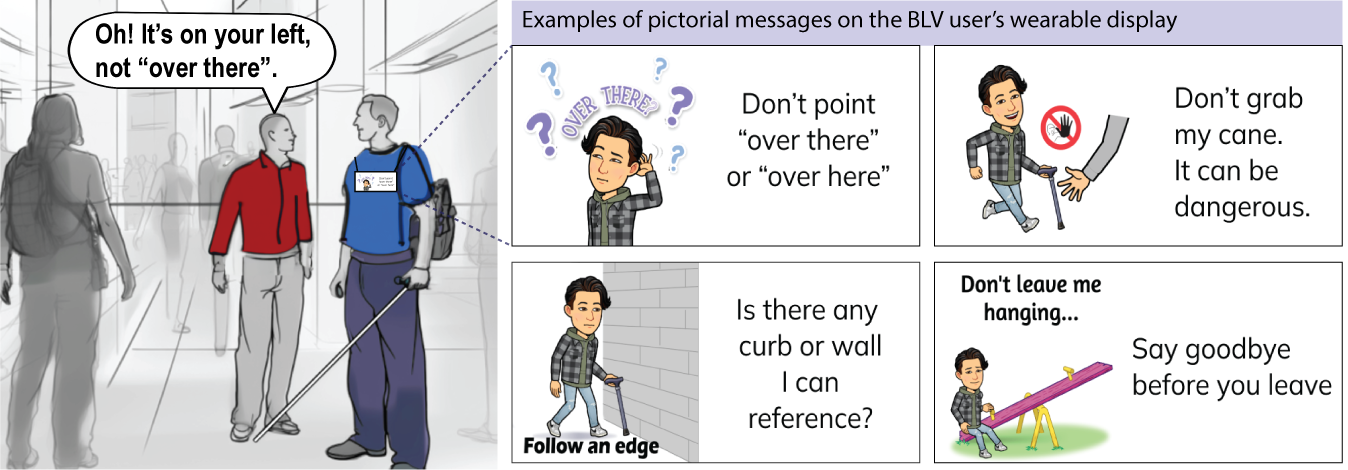}
    \caption{\textbf{\textit{Pictorial Display}}. Left: A BLV user (in blue) wearing the \textit{Pictoral Display} on their chest receiving help from a sighted user (in red). The sighted user upon seeing the message on the \textit{Pictorial Display} makes a correction of their words. Right: Examples from a set of 20 pictorial messages used in our study.}
    \label{fig:system_pictorialdisplay}
    \Description{Two sets of diagrams depicting how the Pictorial Display prototype works. On the left, a diagram depicts a BLV user (in blue) wearing the Pictorial Display screen on their chest while receiving help from a sighted user (in red). The sighted user is shown initially using the phrase “over there”, but upon seeing the message shown on the Pictorial Display, corrects their words to say “on your left” instead. On the right, four examples of pictorial messages that appear on the Pictorial Display are shown. These messages include, “Don’t point ‘over there’ or ‘over here’”, “Don’t grab my cane. It can be dangerous”, “Is there any curb or wall I can reference?”, and “Don’t leave me hanging—say goodbye before you leave.” }
\end{figure*}

\begin{figure}
    \centering
    \includegraphics[width=0.47\textwidth]{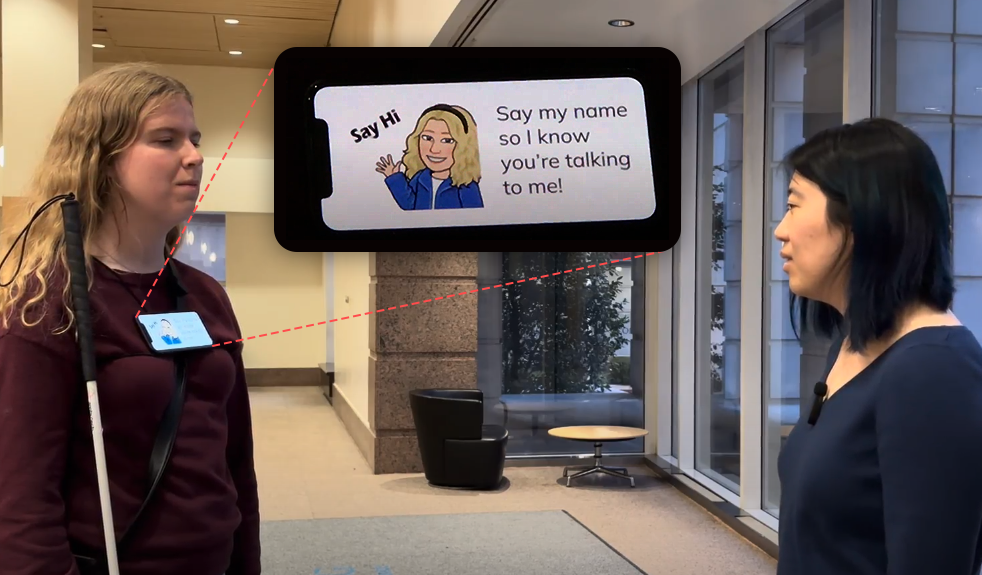}
    \caption{\revision{\textbf{\textit{Pictorial Display implementation.}} \newline The BLV user's \textit{Pictorial Display}, worn high on their chest, reminds the sighted helper to begin by explicitly mentioning the BLV user's name. Appendix~\ref{sec:bitmoji-reference} shows all possible pictorial messages.}}
    \label{implement_3}
    \Description{A photograph shows a blind person with a white cane and a sighted person standing and facing each other, with the blind person wearing a smartphone horizontally across their upper chest. The sighted person notices the display, which has a playful avatar saying “Hi!” and the message “Say my name so I know you’re talking to me!” next to the avatar.}
\vspace{-6pt}
\end{figure}

\begin{figure*}
    \centering
    \includegraphics[width=1\textwidth]{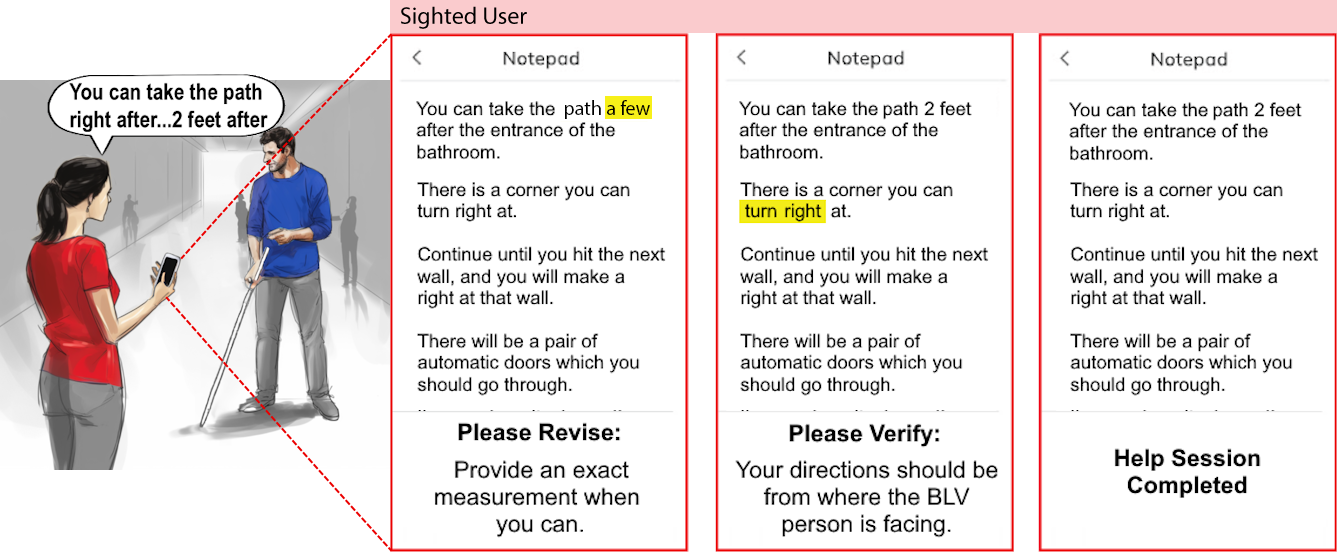}
    \caption{\textbf{\textit{Vague Direction Flagger.}} Left: A BLV user (in blue) receiving help directions from a sighted user (in red) who has the \textit{Vague Direction Flagger} app on their phone. Upon seeing the direction they are giving is flagged as vague by the app, the sighted user provided more specific information. Right: Screenshots of the \textit{Vague Direction Flagger} showing two examples of flagged vague wording and a notice of help completion following corrections of all the vague directions.}
    \label{fig:system_flagger}
    \Description{A diagram on the left depicting how the Vague Direction Flagger prototype works, accompanied by some screenshots of the Vague Direction Flagger app interface on the right. The diagram depicts a BLV user receiving help directions from a sighted user who has the Vague Direction Flagger app on their phone. Upon seeing that the directions they are giving have been flagged as vague by the app, the sighted user provides more specific information. In the screenshots of the app interface, a text transcript of the verbal directions that the sighted user gave is shown. Vague words are highlighted in yellow, and each highlight is accompanied by a short suggestive message, such as “Provide an exact measurement when you can.”}

\end{figure*}

\begin{figure}
    \centering
    \includegraphics[width=0.47\textwidth]{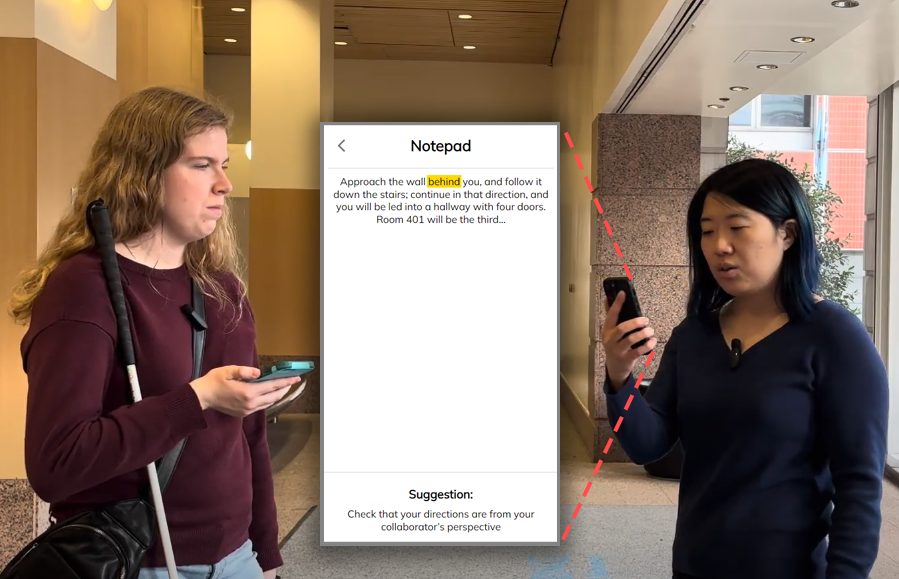}
    \caption{\revision{\textbf{\textit{Vague Directions Flagger implementation.}} \newline As the sighted helper gives instructions, the \textit{Vague Directions Flagger} app on their phone prompts them to check that their directions are oriented from the BLV user's perspective.}}
    \label{implement_4}
    \Description{A photograph shows a blind person with a white cane and a sighted person standing and facing each other, with the sighted person looking at their phone. An inline enlargement of the sighted person’s phone screen shows photograph shows the smartphone app flagging the word “behind” in the instructions, “Approach the wall behind you” as vague, suggesting to the sighted person to “Check that your directions are from your collaborator’s perspective.”}
\end{figure}

\section{Help Supporter Prototypes}
\label{sec:systems}
% We designed our four prototypes through a process of ideation and iteration, two for the Connection Phase and two for the Collaboration Phase.
% During the design process, we first identified the parties involved in the process of help - help supporters, the BLV user who is looking for help, the sighted user who is potentially providing help, and the public environment.
% We then brainstormed the possibilities of how the parties could work together during each phase, what roles they play, and how they interact with each other.  
% This process resulted in ten prototypes. We refined and converged to the two prototypes for each phase that are drastically different and represent the range of diversity. 

We developed four prototypes using a \minor{research-through-design process}, with two prototypes for the connection phase and two for the collaboration phase. 

\minor{Over the course of several months, we designed the prototypes through ideation and iteration. 
Two authors started with an idea of using fun-friendly Bitmoji ~\cite{bitmoji_nodate} public messages to bridge the social differences between sighted and BLV people. Following this initial prototype idea, we engaged in} extensive brainstorming and internal co-design sessions among four mixed-ability co-authors (three sighted and one low-vision author) and another low-vision lab member (mentioned in acknowledgment) who was on a different project team. \minor{During these discussions, the debate between private and public display of messages, as well as information format and level of detail, started to emerge. Our mixed-ability co-designers conducted a series of regular meetings. We created note cards for all the diverse perspectives and considerations from both BLV and sighted team members. Then, we used affinity diagramming to iterate and organize them into groupings. }
% We self-identified as the human stakeholders in the help-seeking process, BLV users in need of assistance and sighted users potentially offering help, who interact with non-human stakeholders, the \textit{Help Supporters} and the surrounding environment. 
\minor{Together, }we identified six design attributes that the co-designers unanimously agreed were important but could be designed in different ways. For example, within what radius should a help supporter connect sighted strangers willing to assist, and should the process of requesting help be discreet or be announced more publicly? We formulate these six design attributes in Table \ref{fig:design_axes}. 
% The design attributes guided our iterative process such that each prototype we could represent divergent values and bring them into our investigation.
% For the Connection Phase, we identified three axes. The first axis is discreetness - to what degree can BLV and sighted people connect for help discreetly without drawing attention to themselves. The second axis is decision-making agency - who can make decisions to connect for help. The third axis is environmental distance coverage - How far away can people connect for help.
% For the Collaboration Phase, we 
%We started our design process by identifying the stakeholders involved in the help-seeking process, which includes the \textit{help supporters}, BLV users in need of assistance, sighted users potentially offering help, and the surrounding environment itself. 
%Subsequently, we conducted brainstorming sessions to explore various possibilities regarding how these stakeholders could work together during each phase, their respective roles, and the nature of their interactions. 

\revision{Next, given these six design attributes, the co-design team employed an iterative ideation process to generate a list of more than ten prototype ideas with various combinations of the target design attributes.}
%This iterative ideation process yielded a total of ten concepts for prototypes. 
\revision{From this pool, we selected two prototype ideas for each phase of help (the connection phase and collaboration phase) and developed them into full prototypes. Table~\ref{fig:design_axes_withvalues} shows the design attributes for the four final prototypes. Table \ref{fig:summary_prototypes} summarizes the four prototypes and how they fit into the interdependence framework proposed by Bennett et al.~\cite{bennett_interdependence_2018, vincenzi_interdependence}; specifically, how the assistive technologies interact not only with disabled people but also sighted people and the environment to foster the shared goal of facilitating access.}  \enlargethispage{-10pt}
% \yy{The back to back mention of Table 2 and Table 3 are still weird.}

% selected two distinct prototypes for each phase that encapsulate a wide spectrum of diversity in terms of how help supporters could be designed.}

%It was during this process we conceived our investigative objectives for each phase.
Our two \textit{connection phase} prototypes explore different approaches for how BLV and sighted strangers could connect with each other. The first prototype, \textit{Person-Finder Glasses}, fosters face-to-face requests for help. It enables the BLV user to locate potential helpers nearby so that they can ask for help face-to-face. The second prototype, \textit{Volunteer Platform}, fosters a mobile-app volunteer platform for requesting help, similar to BeMyEyes but for in-person help. In this approach, help seekers are matched on the app with nearby pre-enrolled sighted volunteers, after which the sighted helper walks over to meet the BLV person in person.

Our two \revision{\textit{collaboration phase}} prototypes explore different ways for the \textit{help supporters} to provide real-time information to the sighted party during the process of helping to improve the assistance they are giving. The \textit{Pictorial Display} prototype explores a public display with a causal and fun-friendly information format. The \textit{Vague Directions Flagger} prototype explores a private display with a formal and detail-specific information format. 

\accept{Below, we describe each prototype and their implementation in detail. }

% Table \ref{fig:summary_prototypes} summarizes how each of our \textit{help supporter} prototypes is designed to support help in the connection and collaboration phases, and how they interact with BLV users, sighted users, and the physical environment. 

\subsection{Connection Phase Prototype 1: \textit{Person-Finder Glasses}}
\label{sys:glasses}

\textit{Person-Finder Glasses} are a HoloLens-based tool intended to provide a direct means for BLV individuals to connect face-to-face with potential sighted helpers. Its design is shown in Figure \ref{fig:system_personfinder}, implementation shown in Figure \ref{implement_1}.
%create an experience for BLV people to directly say hello to potential sighted helpers in person. 

In public environments, BLV people have difficulties locating \accept{other} people in their surroundings and are hesitant to reach out for help due to concerns about behaving in a socially awkward manner. For instance, when asking for help and not knowing surrounding strangers' exact locations, BLV people often resort to loudly shouting out to attract others' attention. 
We designed the \textit{Person-Finder Glasses} to help BLV people locate, approach, and greet nearby strangers. The technique we use is inspired by prior work that leverages facial recognition to help BLV people in social interactions \cite{morrison_peoplelens_2021, lee_pedestrian_2020}. 
Through this prototype, we explore how the \textit{help supporters} may connect people directly face-to-face, and what BLV and sighted users' attitudes are toward this approach.

\textit{Person-Finder Glasses} leverage computer vision technology and provide audio feedback to support BLV users in locating nearby strangers without the need for shouting. When wearing the \textit{Person-Finder Glasses}, BLV users receive a continuous auditory cue when the headset detects a nearby stranger in the direction they are facing. This auditory cue guides the BLV user to walk forward toward the detected stranger. As the BLV user approaches the stranger and reaches a distance of approximately two feet, the continuous auditory cue is interrupted by another distinct sound cue, signaling the BLV user to halt their walking and greet the stranger by saying hello.
\textit{Person-Finder Glasses} act as a \textit{help supporter} for the BLV user, guiding them to locate and approach a nearby stranger comfortably. This approach aims to eliminate awkward behaviors such as shouting or unintentionally bumping into others, thus improving the experience for connecting BLV users and sighted strangers. 

\textit{Person-Finder Glasses} is implemented using Unity on Microsoft HoloLens (1st gen).
The implementation of \textit{Person-Finder Glasses} consists of Microsoft Holographic Face Tracker \cite{microsoft_holographic_2023} and audio feedback integrated through Unity and deployed to Microsoft HoloLens (1st gen) head-mounted display.
%While the HoloLens' front-facing cameras capture video frames of the BLV user's surrounding environment, the system uses the Holographic Face Tracker to detect human faces. If a human face is detected within 10 feet distance, the HoloLens feeds back two types of audio cues, the first type to guide the BLV user to approach a surrounding person and the second type to instruct the BLV user to stop at a 2-feet distance in front of the person in order to avoid bumping into the person. 
The Holographic face tracking detects both front and side views of human faces, but it does not detect the back of a human head.  \enlargethispage{-10pt}

%\textbf{Prototype B}: Sighted People initiate [Figure \ref{fig:enter-label}]
\subsection{Connection Phase Prototype 2: \textit{Volunteer Platform}}
\label{sys:platform}

%The goal of \textit{Volunteer Platform} is to investigate the agency of sighted people connecting with BLV people who are in needs of help.
\textit{Volunteer Platform}is a mobile-app-based prototype that provides an indirect means of connecting BLV users and their potential helpers. Unlike the \textit{Person-Finder Glasses}, which helps the BLV person ask for help face-to-face, the \textit{Volunteer Platform} allows the BLV person to ask for help indirectly via an app. The app matches them with a nearby volunteer, then guides the volunteer to the BLV person's location so the volunteer can help them in person. Figure \ref{fig:system_volunteerplatform} illustrates the \textit{Volunteer Platform} design, and Figure \ref{implement_2} shows its implementation.

BLV users often hesitate to ask for help, and sighted users often do not know when BLV users need help and whether they should offer help. The design of \textit{Volunteer Platform} provides a means of matching BLV users' help requests with nearby potential helpers' offering to help, thereby mitigating the connection phase's social barriers, including awkwardness, a sense of imposing a burden on others, or fear of making a false assumption that a BLV person needs help when they do not. The technique we use in the \textit{Volunteer Platform} mirrors existing peer-to-peer applications and research for community-centered helping \cite{bellotti_peer-to-peer_2014}. 

Both BLV and sighted users use the \textit{Volunteer Platform} mobile app on their respective smartphones. 
The app starts by prompting the sighted user to provide basic information about themselves and confirm their availability to help while they have the app open. The information about themselves includes their first name, how frequently they are in the area (``Regularly,'' ``Sometimes,'' or ``First Time''), and whether they are a staff member or just a visitor in the area. 

BLV users initiate help requests on the app. The help request includes their first name, type of mobility aid, the type of assistance they need (``Physical Guidance,'' ``Directions to a Destination,'' or ``Description of Environment''), and the amount of time they figure they need for the help. Following this, the app generates and displays to the BLV user a list of nearby sighted users who are available to help them.

The BLV user can then choose which helpers from this list they would like to share their help request. The specified sighted users receive the help request, and the match is established whenever the first helper accepts the request. The app will then share the BLV person's location with the matched sighted user so they can approach and greet the BLV person in person.

We implemented the \textit{Volunteer Platform} as a web app using React JS. We used Ngrok \cite{noauthor_ngrok_nodate} to generate unique, temporary URLs for our study participants.

%\subsubsection{\textbf{Collaboration Phase}\\}

% These prototypes focus on mediating the interaction, ensuring that collaboration goes more smoothly and is equally informative and effective. The design requirements of the systems is to communicate the needs of BLV people and providing corrective and/or educational feedback to the sighted collaborator. The two prototypes we designed for this phase are as follows:

%\textbf{Prototype C}: \textit{Pictorial Display} [Figure \ref{fig:prototypec}]
\subsection{Collaboration Phase Prototype 1: \textit{Pictorial Display}}
\label{sys:pictorial}

 %The third prototype, \textit{Pictorial Display}, explores a public display with a causal and fun-friendly information format. 

%The goal of \textit{Pictorial Display} is to examine the user preferences to publicly display suggestive and corrective messages on the wearable device of the BLV user. 

\textit{Pictorial Display} is the first of the two prototypes in the collaboration phase which explores how technology can support the process of a sighted person helping a BLV person after the two parties have met. It represents the approach of displaying help information for the sighted helper using a wide variety of information types. It aims to address two common struggles during the collaboration phase: sighted people struggling to follow BLV people's boundaries and communicating directions in a way that does not depend on vision.

\textit{Pictorial Display} is a wearable public display on the BLV user's chest that displays helpful information in real-time to the sighted helper in a casual, lighthearted tone as shown in Figure \ref{fig:system_pictorialdisplay}. We used a smartphone as the wearable display screen. Figure \ref{implement_3} shows its implementation. 
% The \textit{Pictorial Display} explores an approach to support sighted helpers with easy-to-access and fun-friendly pictorial messages based on their interaction with the BLV user. 
% The \textit{Pictorial Display} prototype consists of a wearable display with a smartphone screen size worn by BLV users, on the front chest location for optimal viewing by nearby helpers in the environment. 

The pictorial messages include a set of 20 pre-designed messages (See \revision{Appendix ~\ref{sec:bitmoji-reference}}) derived from BLV people's language and social etiquette preferences \cite{envision_how_nodate, missouri_how_nodate, wisconsin_health_2010, blind_perkins_1970}. Each message consists of a design with Bitmoji avartar \cite{bitmoji_nodate} and a short text phrase. To facilitate the user study, we generated Bitmoji characters that approximately matched BLV participants' gender and ethnic identities. 
We use the Wizard-of-Oz method to control which message to display in the context of the participants' conversation during the user study. Specifically, a study confederate accompanied the study and remotely controlled the display in real-time in a manner not noticed by the sighted participant.

We implemented the \textit{Pictorial Display} in two parts: a web app for the smartphone wearable display and a control dashboard for the study confederate, both implemented in HTML and JavaScript. 
%A web socket server implemented with Node JS allows experimenters to remotely control the displayed messages through the web dashboard with the Wizard-of-Oz method. 

% \yy{TODO: Attach full set of pictorial messages in the appendix}

%\textbf{Prototype D}: Private Display [Figure \ref{fig:enter-label}]
\subsection{Collaboration Phase Prototype 2: \textit{Vague Directions Flagger}} 
\label{sys:flagger}

The \textit{Vague Directions Flagger} is our second help-supporter prototype for the collaboration phase.
% , representing the approach of giving the sighted helper a private (rather than public) display with a specific information type: detecting vague instructions that they give to the BLV person. 
% The \textit{Vague Directions Flagger} 
It is a mobile app on the sighted user's smartphone that reviews the directions and descriptions they give to the BLV person in real-time, flags vague words and phrases, and instructs the sighted person to make corrections, as illustrated in Figure \ref{fig:system_flagger} and implemented in Figure \ref{implement_4}.
In contrast to the \textit{Pictorial Display}, which provides the information publicly and in a pictorial format, \textit{Vague Directions Flagger} provides the information privately on the sighted person's smartphone in a formal, text-only format. 

During the collaboration, as the sighted helper verbally gives the BLV helpee directions to their destination or descriptions of the environment, \textit{Vague Directions Flagger} transcribes the sighted helper's speech into text and displays it on the app. 
Then, the app processes the text and follows the rules in Appendix C to detect words and phrases that constitute vague directions. It will highlight any such words and phrases in a bright yellow color and display a guideline message (also summarized in Appendix C) to help the sighted user correct or clarify the language. 

As an example, the sighted helper might say: ``The door is a few steps to your left,'' and the \textit{Vague Directions Flagger} app would highlight the phrase ``a few'' on the transcript and display the message ``Please revise: provide an exact count or measurement when you can.'' The sighted helper can then verbally correct themselves by saying, ``The door is 10 steps to your left.'' Finally, when all the flagged words and phrases are resolved by the sighted helper, the \textit{Vague Directions Flagger} confirms that no more vague phrases remain. 
% The system is a mobile-based tool on the sighted user's personal smartphone that transcribes the verbal conversation to text and detects and highlights any language that does not match shared non-visual needs or preferred social etiquette. The sighted users are prompted by the highlighted text and suggestive guidance to improve their descriptions until the system deems the collaboration to be fulfilled and helpful. 

We implemented the \textit{Vague Directions Flagger} as a web app using React JS and deployed it through a cloud server. We used Ngrok \cite{noauthor_ngrok_nodate} to generate unique, temporary URLs for each sighted participant. A study confederate manually performed the speech-to-text transcription during the study.

\section{Methods}
We performed a qualitative study evaluating the four approaches represented by \textit{help supporters} prototypes. 
Our user study was conducted in pairs of BLV and sighted participants, in person, on a university campus.
Here we describe our participants recruitment, study procedure, and data analysis procedure. 

\subsection{Participants}
We recruited a total of 20 participants: 10 BLV participants and 10 sighted participants. We randomly paired them in mixed-ability pairs for the study sessions.
We recruited the BLV participants from a mailing list of individuals who participated in
our group’s previous studies, through our group's ongoing collaboration with a major blind-serving organization, and snowball sampling ~\cite{goodman_snowball_1961}. BLV individuals experience a variety of vision conditions. To ensure that we recruited BLV participants who were suitable for our study, we only included participants who answered ``yes'' to the question ``Does your vision affect your ability to navigate unfamiliar places?'' on our study sign-up Google Form. 
Sighted participants were recruited by posting flyers on a university campus. 
The study was approved by our institution’s IRB.
Table ~\ref{tab:participants} summarizes the participants’ demographics.

% We conducted a study with 20 participants (10 pairs) who tested our prototypes. 
%  \outline{Add how participants are recruited, flyers, through non-profit BLV organizations, through snowballing
% 1 sighted participant and 1 BLV participant make a pair
% IRB approval + compensation}

\subsection{Study Procedure}

Our study consists of a three-step procedure: 
a pre-study interview, experiences using each help supporter prototype, and a post-study interview. 
Each prototype usage experience typically lasted between 15 to 20 minutes. The entire study, from start to finish, lasted approximately 120 minutes. 
To facilitate the study, two experimenters worked collaboratively. They conducted the pre-study and post-experience interviews, as well as prototype onboarding concurrently. One experimenter worked with the BLV participant, while the other simultaneously worked with the sighted participant.
During the help supporters prototype usage experiences, one experimenter assumed the role of a designated study confederate. This experimenter operated a laptop discreetly, outside of the participants' view, to control the Wizard-of-Oz aspects of the study.

% The study includes system testing in pairs and semi-structured interviews. 
% We interviewed participants before and after testing the systems in order to understand their experiences.

%Our study involved three stages: a pre-study interview, a prototype usage session, and a post-study interview. Each study was conducted with a pair of participants, one BLV participant and one sighted participant. 
%We conducted all studies in person at three large public spaces at a university: a lounge/food court, a cafe, and an academic building lobby.

\subsubsection{Pre-Study Interview}

In the pre-study interview, we first collected demographic information from both participants, including age, gender, ethnicity, and level of education.
Then, we asked the BLV participant about their recent experiences in seeking help from sighted strangers. 
During the process, we asked them to recall their frequency of seeking help, comfort levels when requesting help, confidence in communicating their needs, and overall satisfaction with the help they received. 
% For the BLV participants, we then asked about their experiences in seeking help from sighted strangers. 
% Specifically, We asked them to recall their most recent experience, with questions such as  ``how often do you ask for help from a stranger?'', also using Critical Incident Technique ``Could you recall the last time you asked for assistance from a stranger in a public place? How did you find assistance?  How did you ask for/receive assistance? “ how they tend to ask others for help, how comfortable they feel asking for help, how confident they feel communicating their needs with others, and how satisfied they typically are with the help they receive. Collecting this information helped us identify individuals’ prior experiences before testing our prototypes' preliminary preferences for how to connect and collaborate with sighted strangers when they need assistance. 
We asked the sighted participant whether or not they had past experiences interacting with BLV people. 
If their answer was yes, we asked about their experiences and how confident they feel about helping BLV people. 
These questions were designed to enable both sighted and BLV participants to compare and reflect on the differences in their post-study interviews.

\subsubsection{Prototype Usage Experiences}

\revision{For each pair of mixed-ability participants, we asked the BLV participant to seek help and the sighted participant to be a stranger willing to help. We had participants try a connection phase prototype first, followed by a collaboration phase prototype. This sequence was then repeated with the other connection phase and collaboration phase prototypes. 
%then followed by the other Connection Phase prototype and the other Collaboration Phase prototype. 
This is to allow participants to experience the entire sequence of help (both phases) together rather than breaking it apart. 
Participants experienced each phase's prototypes in counterbalanced order. 
Half of the participants started with the Person-Finder Glasses and then tried the Volunteer Platform, while the other half followed the reverse order. The same counterbalancing was applied to the collaboration phase. 
%That is, half of them tried the Person-Finder Glasses first and the Volunteer Platform later and half the other way around, and same for the Collaboration Phase. 
We conducted our study in several locations---two academic building lobbies, a student lounge, and a cafe---employing different locations for different pairs of participants as a way of understanding attitudes and behaviors in diverse environments.}
% The prototype usage experiences of our user study allowed the BLV and sighted participants to try all four prototypes in pairs, with the BLV participant seeking help and the sighted participant as a stranger helper.
% In order to control for order effects, we randomized the order of the prototypes and the order of physical locations where the usage experience took place. The physical locations are two academic building lobbies, a student lounge, and a cafe. 

Before each prototype experience, two experimenters separately but concurrently gave the two participants individualized onboarding instructions, so that they understood how to use the system from their perspective but were not fully aware of how the system supports the other party. The experimenter working with the BLV participant also helped them put on the HoloLens head-mounted device for the \textit{Person-Finder Glasses} prototype and the wearable display phone screen for the \revision{\textit{Pictorial Display}} prototype. The experimenters distributed unique URL links for accessing the \textit{Volunteer Platform} and \textit{Vague Directions Flagger} prototypes. Participants used their own smartphones for these two prototypes. BLV participants used the accessibility feature on their smartphones to access the web app as needed. 

During the usage experiences for the connection phase prototypes, the two experimenters physically separated the two participants to simulate the process of locating and approaching strangers. 
We created six pseudo helper profiles with a variety of basic information on the \textit{Volunteer Platform} (see Section ~\ref{sys:platform}) and instructed the sighted participant to add their information on the platform. We structured this part of the study this way so that BLV participants could experience the process of selecting the helper from a list of nearby potential helpers.

During the usage experiences for the collaboration phase prototypes, one experimenter accompanied both participants while the other experimenter acted as a study confederate by controlling the Wizard-of-Oz aspects of the study. As described in Sections ~\ref{sys:pictorial} and ~\ref{sys:flagger}, the study confederate remotely controlled which pictorial message to display using the dashboard, and transcribed the sighted participant's speech-to-text through manual typing.

The experimenters took observation notes during the study for later analysis.

%The BLV participant selected a few options from a list of helpers, and the sighted participant received a help request notification. The sighted participant accepted the request and approached the BLV participant.

%\textit{Vague Directions Flagger}: Once the participants were connected, we tested the next system. 
%The BLV participant asked the sighted helper a question about the space that we were in (i.e., “Where is the bathroom?”, “Can you describe this space to me?” etc.). As the sighted participant responded, we transcribed the response. The system prompted the sighted participant with feedback on any ambiguous language in the description. The sighted participant corrected any ambiguous language before concluding the help session.  
%We tested the next two prototypes in an open room with little to no obstacles. 
% \textit{Person-Finder Glasses}: Participants stand across the room from each other. The BLV participant uses the audio cues from the system to find the sighted participant and stand within speaking distance. 
%\textit{Pictorial Display}: Once the participants are within speaking distance, the BLV participant asks the sighted participant questions similar to the ones asked for the \textit{Vague Directions Flagger}. The sighted participant is prompted with tips such as “Don’t point”, and “Use exact counts when possible” while speaking to make sure the BLV participant is able to 

\subsubsection{Post-Study Interview}

After the participants experienced all four prototypes, we conducted post-study interviews with the two participants concurrently in separate rooms.
All interviews were recorded with participants' consent and transcribed for analysis.

In the semi-structured interviews, we asked questions including how participants felt about their experiences in the connection phase and the collaboration phase, their preferences for the approaches represented by the help supporter prototypes, the types of information they found useful, additional information they wanted to have, and how their experiences in the study compared to their past experiences getting or giving help. 
% Interviews were based loosely on the pre-prepared questions. 
We asked unscripted follow-up questions to dig deeper into insights raised by participants.

\subsection{Data Analysis}
% \outline{TODO: Describe interview transcription and thematic coding process.}
% \outline{TODO: Describe log data, during the study recordings, observation notes}

%\yy{copy from social wormholes placeholder}

After transcribing the post-study interviews, two researchers independently sectioned the transcripts into quotes for a bottom-up, open-coding approach to data analysis \cite{charmaz_constructing_2006}. We also added the observation notes that we took during the study to the analysis. 
Next, the researchers worked through multiple rounds of meetings to iterate on the codes, discuss their similarities and differences as part of a comparative analysis \cite{merriam2015qualitative}, and leverage them in an affinity diagramming process \cite{holtzblatt2017affinity}. 
The researchers determined that they reached code saturation when neither researcher could identify new codes or arrive at new interpretations of the existing codes after several rounds of revisiting the quotes. 
%In accordance with Mcdonald et al. \cite{mcdonald_irr}, we did not compute inter-rater reliability (IRR), since we used the coding process to discover emergent themes or recurrent topics and permitted multiple possible interpretations of the meaning of the codes. 
%After the two researchers completed their synthesis of an affinity diagram, two additional researchers reviewed the themes and provided their comments. The themes and sub-themes that emerged from this process highlighted ...

\begin{table*}[t]
\caption{Self-reported demographic information of our participants who completed the study. B=BLV participant, S=Sighted participant.\newline \small \normalfont{Note: "Otherwise visually impaired" serves as a general term that describes having some usable vision and may perceive shapes, light, or movements to some extent. However, as the participants indicated in our sign-up Google Form and pre-study interviews, their vision cannot give them sufficient information about the environment such that they face challenges venturing out in public alone.}}

\label{tab:participants}
\tiny
\resizebox{0.8\linewidth}{!}{
\begin{tabular}{lcclll}
\hline 
\textbf{Pair ID} & \textbf{Gender} & \textbf{Age Group} & \textbf{Education Level} & \textbf{Ethnicity} & \textbf{Vision Level}

\\ \hline

P1B  & Male & 18 to 25 & Bachelor's Degree & South Asian & Otherwise visually impaired \\
P1S  & Male & 18 to 25 & Bachelor's Degree & White &  Sighted\\
\hline
P2B & Male & 46 to 55 & PhD & White & No usable vision \\
P2S & Female & 18 to 25 & Bachelor's Degree & South Asian & Sighted\\
\hline
P3B & Male & 26 to 35 & Bachelor's Degree & East Asian & No usable vision\\
P3S & Female & 26 to 35 & Masters Degree & White & Sighted\\
\hline
P4B & Female & 36 to 45 & High School/Some College & East Asian & Otherwise visually impaired\\
P4S & Female & 26 to 35 & Masters Degree & White & Sighted\\
\hline
P5B & Female & 26 to 35 & Bachelor's Degree & East Asian & Otherwise visually impaired \\
P5S & Male & 18 to 25 & High School/Some College & White & Sighted\\
\hline
P6B & Female & 26 to 35 & High School & East Asian & Otherwise visually impaired\\
P6S & Female & 18 to 25 & High School/Some College & White and Asian & Sighted\\
\hline
P7B & Female & 26 to 35 & Masters Degree & White & Otherwise visually impaired\\
P7S & Female & 18 to 25 & Bachelor's Degree & South Asian & Sighted \\
\hline
P8B & Female & 18 to 25 & High School/Some College & White & Otherwise visually impaired\\
P8S & Female & 18 to 25  & Bachelor's Degree & East Asian & Sighted \\
\hline
P9B & Male & 46 to 55 & Masters Degree & African American & Otherwise visually impaired \\
P9S & Male & 18 to 25 & High School/Some College & White & Sighted \\
\hline
P10B & Female & 18 to 25 & Bachelor's Degree & White & Otherwise visually impaired \\
P10S & Male & 18 to 25 & Bachelor's Degree & South Asian & Sighted \\
\hline

\end{tabular}}\

\end{table*}

%%%% 4_v2_methods.tex ends here %%%%

%%%% 5_v4_Findings_Connect.tex starts here %%%%

\section{Findings: Connection Phase}

% In the Connection Phase, our research questions aim to understand how help supporters should help the connection phase. First, How should the \textit{help supporters} encourage in-person requests or foster an app-based platform for requests? 
% Informatio needs = What types of information should \textit{help supporters} facilitate to both BLV and sighted people?

Our research questions related to the connection phase, RQ1 and RQ2, are directed towards understanding the optimal approach for help supporters to bring two strangers together---a sighted individual and a BLV individual---for the purpose of assistance. 
%One option is for the \yy{TODO:  [assistance mediator / AT facilitator / facilatating agent / faciltator / help agent]} to focus on giving the BLV person more confidence in asking a nearby stranger for help face-to-face.
One option, represented by \textit{Person-Finder Glasses}, is for the help supporter to empower BLV person with more confidence to approach nearby strangers to ask for help face-to-face. 
%Another option is for the AT facilitator \yy{replace this term} to connect the other two parties via an app-based volunteering platform similiar to BeMyEyes, avoiding the need for face-to-face requests for help.
Another option, represented by \textit{Volunteer Platform}, is for the help supporter to facilitate help requests and accepts through an app-based volunteering platform, then enables the two parties to meet in person and initiate their help collaboration. 
With RQ1, we asked which of these options is better. RQ2 focuses on identifying the types of information that help supporters should facilitate for both BLV and sighted individuals.

% New paragraph
We found both BLV and sighted participants preferred the on-platform connection approach.
This preference was attributed to the \textit{Volunteer Platform}'s ability to reduce social pressure and cultivate feelings of safety and trust (Section ~\ref{sec:findingsRQ1}). Furthermore, we contribute a collection of information types that help supporters should facilitate on the platform (Section ~\ref{sec-RQ2infoex}).

The findings are summarized in Table \ref{fig:table_findings} for easy reference. 

\begin{table*}
    \centering
    
    \caption{Summary of our findings, which correspond to Sections 5.1--5.2 for the Connection Phase and Sections 6.1--6.2 for the Collaboration Phase.
    }
    \includegraphics[width=0.8\textwidth]{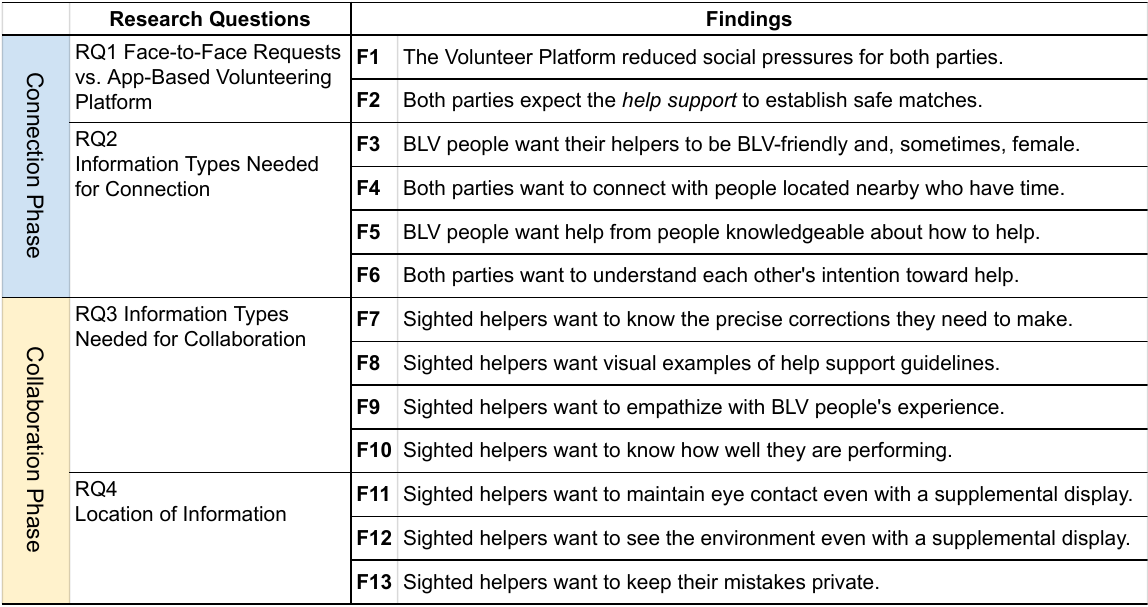}
    \label{fig:table_findings}
    \Description{Provides a summary of our findings for each of our four research questions. These findings are further detailed in Sections 5.1–5.2 for the Connection Phase research questions, and Sections 6.1–6.2 for the Collaboration Phase research questions.}
\end{table*}

\subsection{RQ1 Findings: Face-to-Face Requests vs.\ App-Based Volunteering Platform}
\label{sec:findingsRQ1}
% To understand which approach is better, We examined both in-person and on-platform help requests approach.
Recall that the \textit{Person-Finder Glasses} represented the approach of promoting face-to-face requests, while the \textit{Volunteer Platform} represented the approach of using a mobile app platform for facilitating requests and offers for help.
Our findings indicate that while both approaches can enhance social acceptance and boost confidence for BLV individuals, the app-based \textit{Volunteer Platform} reduces social pressure even more, not only for the BLV participants but also for sighted individuals. As a result, the on-platform approach emerged as our participants' more preferred option. Additionally, both sighted and BLV participants shared a concern for safety and trust, and we found that on-platform approach plays a greater role in alleviating the concern.
% \yy{COMEBACK TO THIS}
% We found that the answer to be nuanced and contextual. The in-person approach empowers BLV people with the new ability to voice their needs with socially acceptable behaviors. Compared to in-person connection, participants preferred peer-to-peer connection using the \textit{Volunteer Platform} because the system platform enables a sense of community, reduces social pressure, and verifies trust and safety among community members. 

% \subsubsection{Social Pressure}
\subsubsection{The \textit{Volunteer Platform} reduced social pressures for both parties.}
%We found that in-person connection approach empowers BLV people to reach out and act in a socially accepted manner to voice their needs. 
%Our study found that \textit{help supporters} have the potential to provide social acceptance, social confidence, and avoid embarrassment when BLV people initiate connections. 

% Our findings suggested that both help supporters faciliated in-person and on-platform approaches improve BLV individuals' social acceptance behaviors and avoid embrassment. 
% BLV participants stated that taking initiative by using the \textit{Person-Finder Glasses} changed their social behavior. Previously, they resorted to desperately and embarrassingly shouting when reaching out. \textit{Person-Finder Glasses} gives them confidence when reaching out for help and behaving in a socially acceptable way.

Our findings indicate that both the direct in-person and on-platform approaches facilitated by help supporters significantly enhance BLV individuals' social acceptance behaviors and effectively mitigate feelings of embarrassment. 
However, it is noteworthy that we also found that only the on-platform approach alleviated sighted individuals' feelings of social pressure.

BLV participants expressed that using \textit{Person-Finder Glasses} to ask for help in person had a transformative impact on their social behavior. Previously, they often found themselves resorting to desperate and awkward shouting when seeking help, resulting in discomfort. Their study experience with \textit{Person-Finder Glasses} provided them newfound confidence when reaching out for assistance, enabling them to adopt more socially acceptable and composed behaviors, as stated by P7B: 

\begin{quote}
\textit{``I think having a way to know with confidence that someone is there and then being able to approach them it's very comfortable. 
I don't like how it makes me feel when I'm shouting in the street because other people might think that maybe I wasn't all there.''} - P7B
\end{quote}

%\textit{help supporters} provide BLV people with great social comfort in reaching out for help, by helping them avoid moments of embarrassment, validating their non-visual senses, and improving the social acceptance of their behaviors in public.

Some participants shared valuable insights on how \textit{Person-Finder Glasses} played a role in validating their non-visual senses, effectively averting them from embarrassment and bolstering their self-assurance. 
BLV participant P3B shared with us that, although he has very good echolocation skills and can hear free-standing entities in space, he often found himself in embarrassing situations asking a nonperson object for help:

% \begin{quote}
% \textit{``It gets embarrassing sometimes. Because, to minimize my embarrassment [of] talking to one person to the next person, I just shout in the general direction. I hear a couple of things because when I make noise when I talk, I can hear large objects, like tables or walls. So if I think I hear something over there, I shout, "Hey, where am I going?" And it turns out what I'm shouting at is something like a garbage can. And then a person standing to my left will be like, "Oh, the street is over here". And I think to myself, "Yeah, okay, I'm just gonna ignore the fact I was talking to a garbage can.''} - P3B
% \end{quote}

\begin{quote}
\textit{``It gets embarrassing sometimes [...] I hear a couple of things because when I make noise when I talk [...] I shout, "Hey, where am I going?" And it turns out what I'm shouting at is a garbage can. And then a person standing to my left will be like, "Oh, the street is over here". And I think to myself, "Yeah, okay, I'm just gonna ignore the fact I was talking to a garbage can.''} - P3B
\end{quote}

% With the tech mediator, they could get confirmation of their echolocation senses before acting upon them, and therefore avoid such embarrassing moments.
% \textit{help supporters} can help BLV people behave in a socially accepted manner and avoid awkwardness. 
% Social acceptance also benefits sighted people when BLV people behave similarly to themselves in a socially familiar manner. 

% Lower barrier to asking for help - 
Regarding the \textit{Volunteer Platform}, we found that the on-platform connection approach was able to lower social pressure for not only the BLV people asking for help but also for the sighted people who were navigating the decision on whether they should accept the help request. 
%We found that on-platform approach lowers social pressure for BLV users to request for help, therefore, increasing their likelihood to ask for help. 
%Without \textit{Volunteer Platform}, BLV users would only ask for help when they become frustrated and absolutely need help. With \textit{Volunteer Platform}, BLV users feel more comfortable and more at ease requesting help. 
%A significant factor in the reduction of the social pressure is that connecting on peer-to-peer platforms enables sighted strangers to silently decline to help, making BLV people feel like they are less of a burden to others.
We found that the on-platform approach effectively reduces social pressure for BLV individuals when seeking help, thereby increasing their likelihood to ask for assistance. In the absence of the \textit{Volunteer Platform}, BLV people tended to resort to seeking help only when they were frustrated and in dire need. The on-platform approach has led to BLV participants feeling more at ease and comfortable when requesting assistance. A contributing factor to reducing social pressure is that engaging on the \textit{Volunteer Platform} enables sighted strangers to silently decline assistance requests. This mechanism fosters an environment where BLV individuals perceive themselves as being less of a burden to others, further promoting a sense of ease and confidence in seeking help.

As participant P8B reflected:

% \begin{quote}
% \textit{"I would feel better about [asking for help]. Because I don't have to feel like I'm putting them in a situation where they can't say no, right? They are on the app, and they accepted the request. So that's good." } - P8B
% \end{quote}

\begin{quote}
\textit{``If they [sighted strangers] don't have time, if they have somewhere to be, they don't have to say no to you. They can just not accept your requests on the app, and you wouldn't even know about this. I think it makes it less awkward for them probably. So I'd feel better.''}  -- P8B
\end{quote}

%%MOVED TO NEXT SUBSECTION IDENTITY INFO
% \subsubsection{Sense of Community}

% On the sense of community, we found that many BLV participants reported that they perceived sighted helpers on \textit{Volunteer Platform} as friendly to the BLV community, willing to help, or having prior training/experience with helping BLV people. This echo previous research on trusting strangers in the shared social group \cite{stolle_trusting_2002}.

% \begin{quote}
% \textit{The app I think is nice because the person if they're on the app, maybe they've already done this before. So I would imagine that would make it better because if they've helped a blind person before, that's probably going to make things a lot better. And then it's also good because when the person wants to help you, it's an easier interaction to have when [it's] already agreed through the app that the person wants to try and help you.} -- P8B
% \end{quote}

\subsubsection{Both parties expect the help supporter to establish safe matches.}
%\subsubsection{Trust and Safety}
\label{sec-RQ1safety}

%\textit{help supporters} have a duty to provide safety information to BLV users.

We found that both BLV and sighted participants shared concerns regarding trust and safety.
Despite the newfound capability of BLV participants to locate nearby individuals, our study revealed that some participants had hesitations in seeking help due to their inherent difficulty in placing trust in strangers.
% As revealed in our study, although the BLV participants enjoyed the new ability to locate surrounding strangers, some of them were still worried about reaching out for help because they struggle to trust strangers. 
%P10B commented when using the \textit{Person-Finder Glasses}: 
\begin{quote}
\textit{``I felt uncomfortable reaching out for help because...you also don't know anything about the person you're approaching. It can be a scary world in [a big c]ity.''} - P10B
\end{quote}

%BLV users indicate they want more information about the other people in the environment. Sighted people often rely on visual to get a sense of trust with strangers before establishing contact. BLV people, due to the lack of vision, seek alternative information to fill the gap, such as gender information and profile pictures. This finding aligns with prior work which reveals that BLV people desire to know whether the stranger is holding a weapon in order to avoid unsafe situations \cite{hurst_is_someone_there}. 
In order to mitigate trust and safety concerns, BLV users expressed the need for additional information about other individuals in their environment. Sighted people often rely on visual cues to establish a sense of trust with strangers before initiating contact. In contrast, BLV individuals, due to their lack of visual perception, seek alternative information sources to bridge this gap.
%This may include details such as profile pictures, which they might enlarge to view or have descriptions before meeting the stranger. 
This finding is consistent with prior research that highlights BLV individuals' interest in discerning whether a stranger is carrying a weapon to avoid potentially unsafe situations \cite{hurst_is_someone_there}.
%When testing \textit{Volunteer Platform}, sighted partcipants initiate by providing their basic identity information, including pseudo-firstnames. 
%Some BLV participants shared that they were able to parse gender information from pseudo-firstnames. P8B, who is a female participant, preferred to get help from female strangers, and she decided to choose only female-sounding names to send help requests to. The possibility of choosing female helpers makes her feel safer. 
% For example, some BLV participants who have limited usable vision preferred to see a profile picture of the stranger enlarged on their phone before connecting with them.

% \begin{quote}
%     \textit{``Depending on the extent of the person's usable vision, it might also be beneficial to have the helper take a selfie or upload a picture of themselves, for security purposes.''} – P5S
% \end{quote}

% \yy{TODO: find a BLV quote}

%Safety concerns is not limited to BLV participants, but also echoed by sighted partcipants. 
%Both sighted and BLV participants expressed a desire for \textit{help supporters} to act as a safety filter through a verification or moderation mechanism.
Our sighted participants also had safety concerns. They expressed a similar unease and emphasized the importance of \textit{help supporters} serving as a safety filter through mechanisms such as verification or moderation, in addition to merely serving as a connector to BLV people who need help:

\begin{quote}
\textit{
%For the connection phase, I prefer [\textit{Volunteer Platform}] over [\textit{Person-Finder Glasses}], because you could potentially have the BLV people indicate and verify that they’re background-checked on the app. 
``Some people, like myself, might not be willing to reach out to any non-background-checked stranger, especially at night. Since neither the BLV people nor the sighted helper knows each other, both might be otherwise uncertain if they can trust the other person. But this problem is eliminated when each of us does the verification on the app.''} - P2S
\end{quote}

% \yy{Design Implication: \textit{help supporters} may play a role as a neutral community moderator that establishes order for in-person community. }

%%%%%%%%%%%%%%%%%RQ2%%%%%%%%%%%%%%%%%%%%%%%%%%%%%
\subsection{RQ2 Findings: Information Types Needed for Connection}
\label{sec-RQ2infoex}

%help supporters play a conversational mediation role to support the alignment of expectations and preferences between BLV and sighted individuals a various forms of information exchange. 

% When initiating the Connection Phase, our study found that BLV and sighted participants rely on a set of information exchange to make decisions. The information exchange is essential for them to let others know about themselves and their preferences, as well as to learn about their potential collaboration partners before connecting. 
% We report on the types of information exchanges occurred and desired during our study, both on the providing and receiving ends by BLV and sighted participants.

RQ2 aims to identify the information that help supporters should give the two parties about each other to help them decide if they want to connect. 
% types of information that help supporters must facilitate for both BLV and sighted individuals who seek to connect for help. 
% We found that both BLV and sighted users depend on a specific set of information exchanges to make informed decisions prior to establishing a connection. 
We identify four types of information that AT facilitators should offer the two parties: identity, availability, knowledge level, and intention.

%\subsubsection{Identity~\\}
\subsubsection{Identity: BLV people want their helpers to be BLV-friendly and, sometimes, female.}

% BLV's safety concerns due to lack of info about surrounding strangers, \textit{help supporters} can fill this gap
%(1) identity information of sighted helpers - e.g. gender, employment

% BLV and sighted participants expressed preferences regarding the identity and demographic information of the other party in the helping connection phase. Based on use cases and contextual factors, these can include gender and employment related identity information. 
% %They and appreciated help agent prototypes that enabled the conveyance of such information. 

% Gender identity and community identity

% In our user study, \textit{Volunteer Platform} was evaluated as an approach enabling the exchange of first names between BLV and sighted participants. 
The \textit{Volunteer Platform} probe exchanges members' first names.
Our findings revealed that, while participants welcomed the use of first names, they also expressed a desire for additional identity-related information based on their interpretations using the platform. We found that community identity and gender identity information hold particular significance.

%We found that participants valued the notion of being a part of a BLV friendly community. 
%When testing \textit{Volunteer Platform}, some participants perceived the presence on the mobile platform as belonging to a community. BLV participants explained that, by virtue of being on the mobile app, sighted strangers were implicitly conveying that they were BLV-friendly, and were more likely to have had prior experience helping or interacting with BLV individuals.

Our findings highlighted the value participants attributed to the concept of being a member of a BLV-friendly community. During the evaluation of \textit{Volunteer Platform}, participants recognized their presence on the mobile platform as indicative of a community affiliation. 
BLV participants elucidated that the sighted individuals' presence on the platforms conveyed a message of BLV friendliness.
The information of sighted people's willingness to sign up for volunteer on the mobile platform had a positive impact on developing a sense of social community. 
%Sighted people volunteer has a positive impact on the sense of social community. 
The value of forming a community is echoed by the success of the remote assistance platform BeMyEyes \cite{be_my_eyes}. 

% Furthermore, our study found that BLV participants perceived the volunteer helpers on the platform as being not only friendly towards the BLV community, but also as individuals prepared and experienced to help:

% \begin{quote}
% \textit{``I think people [on the app] would be more willing to help. And maybe also a little more prepared.''} -- P4B
% \end{quote}

% \begin{quote}
% \textit{“If a sighted person is on the app, maybe they've already helped before. So I imagine that that would make [the help experience] better.”} -- P8B
% \end{quote}

% \begin{quote}
% \textit{``[I]f they're on the app, maybe they've already done this before. So I would imagine that would make it better because if they've helped a blind person before...When the person wants to help you, it's an easier interaction to have when [it's] already agreed through the app that the person wants to try and help you.''} -- P8B
% \end{quote}

\begin{quote}
\textit{``[I]f they're on the app, maybe they've already done this before [...] It's an easier interaction to have when [it's] already agreed through the app that the person wants to try and help you.''} -- P8B
\end{quote}

Sighted participants echoed the sense of community, stating that being on the platform validated them as receptive helpers to BLV individuals.

\begin{quote}
\textit{``[Sighted] users register for [the \textit{Volunteer Platform}] because they're open to helping BLV people. But if [the BLV person is just] approaching any stranger, there's no guarantee that the stranger would be receptive to help. So if both the BLV helpee and the sighted helper are on the [platform], I feel like that this would help the visually impaired person already know that they are a receptive person to ask.''} –P9S
\end{quote}

% Some female BLV participants stated that they preferred to ask help from a female sighted stranger. In our study, although no gender information was explicitly given, some participants were able to parse gender information from first names on \textit{Volunteer Platform} where sighted participants took initiative to provide that information. 
% For example, a BLV participant, who identified as a female, mentioned:

% Additionally, we found that both BLV and sighted participants stated a preference for \textit{help supporters} to disclose users' gender.
% We found that there are two reasons for wanting gender identity information. First, some participants associate female strangers with a stronger sense of public trust. Second, there are some helping tasks that is only socially acceptable to ask for help from strangers in the same gender group, such as help finding or using a restroom.
Additionally, we found that several female BLV participants indicated a preference for seeking help from sighted individuals of the same gender. In our study, even though explicit gender information was not provided, some BLV participants managed to deduce gender information from the first names presented on \textit{Volunteer Platform} platform. For instance, a BLV participant, who identified as a female, stated:

\begin{quote}
\textit{``I liked the [platform] because the fact that you can see the [helpers'] names is nice. Because [that makes it easier] if I want to ask a woman for help.''}  -- P8B
\end{quote}

\subsubsection{Availability: Both parties want to connect with nearby people who have time.}

%(1) location; (2) time; (3) knowing help is on the way
% The second type of information we found is useful is availability, which includes time and location availablity. Prior research [CITE] revealed that one major reason for social barrier in connnecting for help is the misaligned expectations bewteen BLV and sighted people. Our sutdy furthered the understanding by providing availability information, 
% One social barrier in connecting for help is the misaligned expectations between BLV and sighted people. In our study, we found that both BLV and sighted participants benefit from time and location availability information exchange.

The second type of information that we identified as valuable is availability, which encompasses both temporal and spatial availability. A significant factor contributing to social barriers in seeking assistance is the misaligned expectations between BLV and sighted individuals. Our study reveals the availability information needed by both parties.

%Time availability information exchange 
%[TODO] The information exchange around time facilitates the communication between BLV people's estimated time to complete the help task and sighted people's expectation to spare their time.
% Both BLV and sighted participants mentioned that it was important for the help agent to convey what tasks the BLV needed help with, as well as the BLV’s estimate of how long the tasks might take. 
%Participants highlighted how this information could help avoid awkward misalignment in the event of a BLV individual needing help for a longer period of time than the sighted helper could spare, or in the event of a BLV individual approaching a sighted stranger for help when the sighted person was preoccupied.
%%%% are there cases sighted people overestimate time commitement, know it doesn't take that much time to help acutally make them more willing to offer help. 

We found that the exchange of time availability information could effectively avert uncomfortable situations and enhance communication between BLV and sighted individuals. It helps to align BLV helpees' estimated time for completing a task with the sighted helpers' free time they can allocate.
%It facilitates in aligning BLV helpees' estimations of the time required to complete a task with the expectations of sighted helpers regarding the time they can allocate.

Participant P10S described how availability informed them to make decisions: 

% \begin{quote}
% “If they [sighted strangers] don't have time--if they have somewhere to be--they don't have to say no to you. They can just not accept your requests on the app, and you wouldn't even know about this. I think it makes it less awkward for them probably. So I'd feel better [about that].  -- P8B
% \end{quote}

\begin{quote}
\textit{``It lets me know what type of tasks the BLV needs help with, as well as how long of a time commitment the task requires. This helps me judge whether I'm able to help with the task or not.''}  -- P10S
\end{quote}

%Some participants shared that self-reported estimated time was often inaccurate and that knowing specific tasks might help better align expectations. 

Moreover, some participants pointed out that relying solely on self-reported estimated time often led to inaccuracies. There is a need for a more effective approach to provide specific details about the tasks at hand, which could lead to a better alignment of time expectations.

% P8S went even further in emphasizing the importance of the help agent conveying specific details about the task(s), asserting that:

% \begin{quote}
%     \textit{``In the app, it asks: `For how long does [the BLV person] need help?' She typed just five minutes, but it might take more than five minutes, right? And so maybe we can also have information about where [the BLV person] wants to go...then I know specifically where I need to get the person to, which can definitely help me.''} -- P8S
% \end{quote}
% “In the app, it asks: "For how long does [the BLV person] need help? She typed just five minutes, but [in actuality] it might take more than five minutes, right? And so maybe we can also have information about where [the BLV person] wants to go. For example, in the building, if she wants to use the bathroom, she can indicate that via the app. And if that's the case, then I know specifically where I need to get the person to, which can definitely help me [gauge the amount of time it might take].”

% In addition to temporal availablity, we found that spatial availability information provides BLV participants two aspects of important information, making them feel confident and supported - First, help is available in the vicinity; Second, help is on the way. 

Besides temporal availability, we found that spatial availability information gave our BLV participants a sense of confidence and support. It assures them that assistance is available nearby and that the helper is coming for them.
% \begin{quote}
%     \textit{``If I'm wearing the headset [\textit{Person-Finder Glasses}], there might not be somebody readily available in the headset’s field-of-detection that I can talk to. So the app [\textit{Volunteer Platform}] would be more expedient in that sense.''} -P9B
% \end{quote}
% Furthermore, BLV users desired that the help agent convey how far away their sighted helper was, in order to gauge 
%Additionally, location availability information conveys to BLV individuals that help is on the way and estimated distance for their helper to reach them. BLV participants find assurance in having the information that their request for help was reaching surrounding people and help is on the way. 
% BLV users also appreciated the assurance that help agent prototypes like the app offered in letting them know for certain that their request for help was reaching other people. 
% Participants compared past experiences of asking fruitlessly for help, only to realize that they were either alone or had been talking to non-living objects (whom they had mistaken for people) in their surroundings:
BLV participants found it more reassuring compared to their past experience of asking for help with uncertainty: 

\begin{quote}
\textit{“Knowing the helper’s name and that help is on the way, is a huge confidence boost already. Versus if I'm by myself without this app, I'm just calling out to any person-shaped objects. Like, it could be a chair, a pillar, or a garbage can instead.”} -- P3B
\end{quote}

% P2B mentioned, 
% “A feature that could be useful is to have something on the app that indicates how far away the person is.” 

% \begin{quote}
% \textit{“I [would like to know] how far away the sighted helper is when you are selecting them; otherwise, I don't know how long it'll take for them to get to me.”} -- P3B
% \end{quote}

\subsubsection{Knowledge Level: BLV people want help from people knowledgeable about how to help.}

% (1) the sighted helping party have the knowledge to fulfill the question asked by the help-seeking party. 
% (2) the sighted helping party understand BLV people's special needs and preferences. 
% (3) BLV people disclose their disabilities and abilities, such that sighted people would know how to help.

The third type of information that we identified as important during the connection phase is the background knowledge of potential sighted helpers, which has two facets: knowledge regarding the specific task for which assistance is requested, and knowledge concerning the unique needs of the BLV helpee.

%We found that there are two types of knowledge information exchange needed between BLV and sighted participants, knowledge about the helping task and knowledge about BLV people's special needs and preferences. 

We found that knowledge of the tasks requiring assistance communicates the mutual understanding of both parties' familiarity with the particular task. This includes gauging the extent of a sighted helper's knowledge regarding the task and the additional information that a BLV individual is seeking in order to accomplish the task.

Within the context of navigation tasks, both sighted and BLV participants emphasized the significance of comprehending their partner's familiarity with the surroundings. 
%BLV participants stated that the information empowers them to select a sighted helper who is well-acquainted with their location, thus enabling efficient assistance in giving them directions:

\begin{quote}
\textit{``I picked a sighted helper who worked in our location, because these people are more familiar with the environment. If I picked someone who was a first-time visitor, they would be as lost as I am. But if I asked a person familiar with the environment, they can easily tell me where to go.''} -- P6B
\end{quote}

% Not only BLV people needs the information about sighted helpers' knowledge, conversely, sighted people wants to know how much BLV people is already knowledgable about the environemtn, such that they can tailor the level of details in their descriptions.
% Sighted participants use the BLV people's familiarity information to leverage the amount of details to provide in their descriptions.

%We found that the need for knowledge information extends beyond just BLV individuals. 
Sighted individuals also seek to understand the extent of a BLV person's familiarity with the location. This knowledge allows them to customize the level of detail in their descriptions accordingly.

% \begin{quote}
% \textit{``I think an additional helpful piece of info is how familiar they already are with the location/space they need help navigating. This way, I have a better idea of what details to leave in and leave out when assisting.''} -- P10S
% \end{quote}

Another background knowledge is the unique needs and preferences of BLV individuals. 
With this information, sighted helpers can better understand the nature of assistance needed by BLV individuals and how to help them. 
BLV individuals experience a diverse range of visual conditions spanning from complete blindness to low vision and encompasses difficulties such as light sensitivity, depth perception, and field of vision limitations. 
% there exists a complex spectrum of abilities and disabilities. Some BLV individuals may even possess a combination of these conditions.
Initiating an exchange of knowledge regarding the needs and preferences of both parties in advance serves as a crucial step. It allows both the BLV individual and the sighted helper for an assessment of the potential effectiveness of collaboration, ensuring a more informed approach to tackling the task together.

%as well as what particular skills or capabilities they might need to have in order to effectively help the BLV people.

\begin{quote}
\textit{``I think it would be helpful for [BLV users] to also provide their level of visual impairment, since different people have different levels of impairment. Also, some people like me may have other physical impairments as well, that require additional other assistance.''} -- P9B
\end{quote}

% Several BLV participants expressed a desire to communicate not only their disabilities but also their abilities, creating a more holistic understanding of their needs and capabilities. 
% This approach could potentially lower the social barrier for sighted strangers when considering help requests. 
% By conveying their capabilities, BLV individuals aim to enable confidence in sighted helpers, facilitating a more receptive response to their help requests.

% \begin{quote}
% \textit{``I would convey via the app that I have some vision, as opposed to having no vision. I feel like people get less intimidated by someone who's not totally blind. It's kind of more familiar territory than, `Oh goodness, you're blind!' Some people get very, very flustered and don't know how to help in that case.''} -- P7B
% \end{quote}

Sighted participants stated that having knowledge about BLV individuals' \revision{functional} needs would assist in sidestepping potential awkward situations. This knowledge could prevent the inadvertent provision of excessive unwanted assistance or, conversely, the unintentional withholding of adequate support to the BLV individual. 

\begin{quote}
\textit{``It would be great for [help supporters] to provide some details about how severe the BLV person’s visual impairment is, so that I know how much help to give. This way, you're not patronizing the BLV person; if they are able to get up and move around on their own, then you know that they don't need my help for that, but I can assist with reading something far away or pointing them in the direction of something. So to sum up, I guess it would help me assess how much help they need.''} -- P4S
\end{quote}

\subsubsection{Intention: Both parties want to understand each other's intention toward help.}
% Receptiveness to Asssitance Partnership

% (1) Sighted people's willingness to help; 

% (2) BLV people's intention to receive help, or their intention to figure it out by themselves (not needing help)

The fourth type of information that we identified is information on people's intentions. Our findings indicate that it is essential for not only BLV people to gauge sighted people's willingness to offer help, but also for sighted people to understand the intentions of BLV individuals---whether they intend to seek help from others or to independently tackle the task without assistance.

% \textit{help supporters} support the exchange of intention on whether or not a sighted person is willing to provide help, and whether or not a BLV person is in need of help. 

% Both sighted and BLV participants mentioned that they appreciated how the use of help supporters as an implicit conveyance of the sighted person’s willingness to help BLV individuals.

% \begin{quote}
% “I felt comfortable requesting assistance via the app, since the other side [i.e. the sighted helper] is predisposed to helping [since they're on the app].” –P10B    
% \end{quote}

In the context of asking for help, participant P1B reflected on their experience and how the \textit{Volunteer Platfrom} clarified both parties' intentions: 

\begin{quote}
\textit{``Reaching out to a stranger for help is difficult because you worry that you'll get a negative reaction from them. You might overwhelm or overstep their boundaries.
The purpose of [help supporters] is clear: both parties know what they're in for. This means that you can cut the hesitation around asking the people on the [\textit{Volunteer Platform}] for help. 
% Essentially, with the app, both the BLV user and the sighted helper know why they're on the app, the BLV user to get help and the sighted user to offer help, so this helps eliminate any hesitation I would have in asking those helpers for help.''
} –P1B  
\end{quote}

% \begin{quote}
% “Reaching out to a stranger for help is difficult, because you worry that you'll get a negative reaction from them. And like, you might overwhelm or overstep their boundaries or whatever; there's a lot of tiny risks. Basically, when physically asking someone for help verbally, versus discreetly asking for help via an app, where the purpose of the app is clear—both parties know what they're in for, since the app has a single defined purpose—this means that you can cut the hesitation around asking the people on the app for help. Essentially, with the app, both the BLV user and the sighted helper know why they're on the app—the BLV user to get help and the sighted user to offer help, so this helps eliminate any hesitation I would have in asking those helpers for help.” –P1B    
% \end{quote}

% In contrast, both sighted and BLV participants also valued the implicit information conveyed, through the help supporters, in indicating whether the BLV party is in need of help or wants to figure things out themselves. 
% For example, one BLV participant remarked:

Conversely, in the context of unwanted help, BLV participants valued the implicit information conveyed through \textit{help supporters}. It could facilitate communicating their preference for not receiving help unless specifically requested, demonstrating that they are discerning about seeking assistance on their own terms.

\begin{quote}
\textit{``I dislike it when sighted people offer unsolicited help, such as just giving you directions or giving you help without you asking for it. Use of [\textit{Person-Finder Glasses}] allows me to effectively convey to others that, if I desire help, I’m capable of reaching out, but otherwise, I’m fine.'' }–P1B   
\end{quote}

% Similarly, sighted participants mentioned how knowing that BLV individuals could request desired help at any time via the help agent helped them avoid unintentionally patronizing the BLV individuals:

We found that this insight is shared by sighted participants, who valued knowing that the BLV individuals have ability to request assistance via \textit{help supporters} enabled them to avoid unintentionally condescending the BLV individuals.

\begin{quote}
\textit{``[BLV people]'s use of the help [supporter] makes me much more confident in going up to them and offering assistance, because I know they have specifically requested and need help. Whereas without a help [supporter], sometimes it's a guessing game of whether the BLV actually needs assistance or not. Sometimes it's more uncomfortable to go up to [a BLV person] in that way [to offer unsolicited help]. As such, I think the help [supporter] improves my confidence in going up to someone, knowing that they specifically requested help.''} –P5S
\end{quote}

% \begin{quote}
% ``BLVs’ use of the help agent makes me much more confident in going up to them and offering assistance, because I know they have specifically requested and need help. Whereas without a help agent, sometimes it's a guessing game of whether the BLV actually needs assistance or not. Sometimes it's more uncomfortable to go up to someone [a BLV] in that way [i.e. offering unsolicited help]. As such, I think the help agent improves my confidence in going up to someone, knowing that they specifically requested help.'' –P5S   
% \end{quote}

% \outline{TODO: expand the writing. add quotes}

% \underline{discreetness}

% (1) Low discreetness - \textit{Person-Finder Glasses} \& C

% (2) High discreetness - \textit{Volunteer Platform} \& D

% \outline{TODO: expand the writing. add quotes}

% \yy{RQ2 is in the connection phase, but 5.2.2 Form Factors have strong overlap with RQ3 collaboration phase too. How do we organize the content?}

%%%% 5_v4_Findings_Connect.tex ends here %%%%

%%%% 6_v3_Findings_Collab.tex starts here %%%%

\section{Findings: Collaboration Phase}

Here we report our findings for the research questions related to the collaboration phase, which were to uncover the specific types of information that help supporters should provide (RQ3, corresponding to Section~\ref{sec:rq3-findings}) during the collaboration and where that information should be situated (RQ4, corresponding to Section~\ref{sec:rq4-findings}). ~\revision{In Appendix ~\ref{sec:autonomy-findings}, w}e also report findings on how help supporters' advocacy for BLV users affected their own feelings of autonomy.

% (~\ref{sec:autonomy-findings}).
% In our user study, we examined a diverse range of messages presented by help supporters to assist sighted helpers in offering effective directions and descriptions to BLV participants. 
% to sighted helpers to improve their 
% to actively support them in providing useful directions and descriptions to BLV participants. 
% These messages were presented in both text-based and pictorial formats, and they were situated both publicly on BLV helpee agent or privately on sighted helper agent. 
% Drawing from our observations and sighted participants' interview quotes, we report on our findings concerning the types of information and the strategic location of the information.
% We anticipate that our findings will illuminate the design of future \textit{help supporters}, enhancing their role in improving collaboration between sighted and BLV individuals.

As we described in Section \ref{sec:systems}, we designed our collaboration phase prototypes to display just-in-time information to the sighted helper to support them in following best practices. They advocated on behalf of the BLV person but did not display information to the BLV person.
% These prototypes did not communicate with the BLV person much, but they did lift the burden of ``teaching'' from the BLV person.
% As a result, although we interviewed both BLV and sighted participants, the representative quotes here all come from sighted participants.
As a result, our findings for these two research questions come mainly from sighted participants.
% We should note that it was mostly our sighted participants who [viewed and interacted with / benefited from] the assistive technology during the collaboration phase. 
% In the Collaboration Phase, the information provided by \textit{help supporters} predominantly caters to the needs of sighted helpers, consequently leading to indirect benefits for BLV individuals. As a result, it is the preferences articulated by sighted participants that carry the greatest significance during Collaboration Phasae. Our inquiry and synthesis are centered around the feedback provided by sighted participants.
% A preface of our finding is that, for Collaboration Phase, the information from \textit{help supporters} is directly serving the sighted person and then the blind person indirectly benefits from it. 
% Therefore, it is the sighted participants' preferences that provide the most insights for the collaboration phase.
% % are dependent on this information.
% We close our findings (Section 6.3) with BLV's perspectives on having \textit{help supporters} advocate for them. 
The findings are summarized in Table \ref{fig:table_findings} for easy reference.

\subsection{RQ3 Findings: Information Types Needed for Collaboration}
\label{sec:rq3-findings}

% Between \textit{Pictorial Display} and \textit{Vague Directions Flagger}, our study examined 20 pictorial messages and 6 categories of highlighted corrective messages.
% During user study, sighted participants described how and why the information helped them to improve their collaboration with BLV particpants. 
% We found that, the information can be designed to support four aspects of the collaboration: 
% first, nudge sighted helpers to think like a BLV person with empathy;
% second, ask sighted helpers to provide specific details in their descriptions;
% third, use pictorial format to provide examples and humor. 

RQ3 is centered on determining the types of information that help supporters should provide to sighted helpers during the collaboration phase. By conducting usage experiences on both the\textit{Pictorial Display} and \textit{Vague Directions Flagger}, our study observed and interviewed participants to discern their preferences around their usage engaging with 20 pictorial messages (see  Supplementary Material) and six distinct categories of highlighted transcripts flaggers (see \revision{Appendix C}).
We found that sighted participants elaborated on how the information types contributed to their enhancement of collaboration with BLV participants. 
The findings of our study emphasize the significant benefits of these four information types in improving collaboration: specific corrections, illustrations, empathy-building information, and validation.

\subsubsection{Specific Corrections: Sighted helpers want to know the precise corrections they need to make. }

Our findings indicate that sighted participants valued correction information provided by help supporters. In the user study, sighted participants initially gave information that may exhibit vagueness or lack of specificity. \textit{Vague Directions Flagger} reviews the transcripts and subsequently prompts the sighted participants to enhance and elaborate on the information initially provided, thereby augmenting it with more useful details to BLV helpees.

% Another type of information that sighted participants find very help points out the specific language sighted people need to improve and ask them to provide specific details in their descriptions. 

We found that sighted participants liked the highlighting of specific words and phrases in their conversational transcripts. They particularly valued the messages that asked for specific corrections from them. The correction requests help to enrich information and streamline the process of making modifications.

\begin{quote}
\textit{``I think having the suggestions in relation to a specific highlighted area of the directions that needed improvement was more beneficial for me in identifying specifically where I could improve and where my words might have been unclear.'' }-P5S
\end{quote}

% The system plays a role in asking sighted helpers to enrich information in their descriptions, making the collaboration more accurate and informative to BLV helpees.

The type of messages asking for specific details are particularly favored by sighted participants.

\begin{quote}
\textit{``One example of a hint that was really helpful for both [\textit{Pictorial Display}] and [\textit{Vague Directions Flagger}] was the hint about being specific with numbers. So instead of saying "a few steps," we were reminded to say, for instance, "three steps" when specifying distance.'' }-P9S   
\end{quote}

We found that the corrective messages by help supporters have an effect on sighted participants' approach in providing descriptions. Initially, sighted participants tend to use brief, simplistic vocabulary. After seeing messages from help supporters, they started to speak with more comprehensive descriptions with detailed directions, distances, slope degrees, surface types, etc. 
This shift in behavior demonstrates a substantial improvement in their help performance.

% \begin{quote}
% “[\textit{Vague Directions Flagger}] definitely asked us to be very specific about how to [give directions]. For the first direction, I gave a one-word instruction like "straight" or "left", while for the second direction, I actually said a few more details, like “There's a 30-degree ramp". [\textit{Vague Directions Flagger}] requires the instructions to be very specific, so that the BLV user does not get in trouble or fall. I like how the app gives you guidance about which parts [of your directions] need to be more specific.” - P8S
% \end{quote}

\begin{quote}
\textit{``For the first direction, I gave a one-word instruction like  `straight' or `left', while for the second direction, I actually said a few more details, like ``There's a 30-degree ramp''. [\textit{Vague Directions Flagger}] requires the instructions to be very specific, so that the BLV user does not get in trouble or fall. I like how the app gives you guidance about which parts [of your directions] need to be more specific.''}- P8S
\end{quote}

\subsubsection{Illustrations: Sighted helpers want visual examples of help support guidelines.}

In order to address the research question on the types of information that help supporters should provide, we explored two different formats: formal text format and informal illustration format. Our prototype \textit{Pictorial Display} featured an illustration information format, wherein each simple suggestive phrase was complemented by a fun-friendly Bitmoji picture. 
We found that sighted participants preferred this pictorial format and revealed two insights for their preference: first, due to its role as a visual exemplar, and second, as a source of engaging and affable humor.

% In Prototype C, we examined a pictoial information format, where each text-based suggestion is accompanies with a Bitmoji picture. Participants preferred the pictorial information format for two reasons, as visual example and as a source for fun friendly humor.

%% Pictures as Examples
Sighted participants perceived the Bitmoji pictures as visual examples of the suggestions made by \textit{help supporters}. 
The visual examples made it more straightforward and expeditious for the sighted helpers to understand how to adopt the suggestions.
%Participants favored this format due to its capacity for facilitating more straightforward and expeditious comprehension.

% “The image of walls that reminded [sighted helpers] to use those as a kind of reference--I thought that was very, very helpful. Because it's something that I hadn't really thought about. And I'm sure that most people would also benefit from the tips on making use of numerical information about things like stairs, or even reminding people that there are stairs in the first place.” - P6S

\begin{quote}
  \textit{``The pictures [on \textit{Pictorial Display}] gave me examples. For instance, when it said "include reference points", the display had a cartoon of a guy hitting a wall. This gave me the idea: I can think of curves such as the walls [as reference points]. Thus, I think it's more helpful to have these pictures alongside the hints.''} -P9S  
\end{quote}

% \begin{quote}
% \textit{``[\textit{Pictorial Display}] was really helpful: the pictures are really helpful; the sentences were simple, so I could follow it in real time, so that was helpful. ''
% % I also appreciated that the device was [physically] on him, so that you know, I could look at him and still talk to him at the same time, because I feel like I've been told that it's good practice to to look at the person, even if they can't necessarily [see] you. 
% %With [\textit{Vague Directions Flagger}], the hints were still helpful, but it was hard because I didn't want to look at my phone while talking to [the BLV user], since doing so would feel weird
% } - P9S
% \end{quote}

%% Pictures for Humor.
Furthermore, we found that the pictorial information served as a catalyst for humor during the collaboration phase, fostering an enjoyable and amicable atmosphere between sighted helpers and BLV helpees.

During our study, two experiment facilitators observed and noted instances of lighthearted humorous interactions stemming from the use of pictorial information format. 

We observed at one instance, when the wearable screen in \textit{Pictorial Display} displayed a confused Bitmoji character with the text ``Over There'' above its head, the sighted participant reacted with a sudden realization prompting them to exclaim "Oh! Not over there!". This prompted laughter from the BLV participant, who responded with: ``That's a good one! I can't see you pointing!''. Their shared amusement effectively dissolved the momentary sense of embarrassment the sighted participant had felt for their error.

% At the end of a help session between participant pair X, we observed a moment of humorous goodbye. 
% \textit{Pictorial Display} featured a pictorial message of a lonely individual seated on one end of a seesaw with text "Say Goodbye. Don't leave me hanging...". The sighted participant described the picture to the BLV participant before parting ways. They shared laughter concluding the collaboration session on a humorous note. 

Upon being introduced to the pictorial information format by researchers, BLV participants P1B and P7B expressed their endorsement of the design.
%They appreciated the incorporation of jovial and friendly Bitmoji pictures, which contributed to their perceived approachability and engagement with sighted helpers.

% \begin{quote}
%     \textit{``The cartoons on the \textit{Pictorial Display} create a lively and less serious environment.''} -P1B
% \end{quote}

\begin{quote}
    \textit{``The Bitmojis are cool. I think a lot of people will be in favor of that idea... whatever is more engaging is probably going to work best for me.''} -P7B
\end{quote}

\subsubsection{Empathy-Building Information: Sighted helpers want to empathize with BLV people's experience \revision{in order to give better help}.} 
%F9: Sighted helpers want to empathize with BLV people's experience

The third type of information emerged from participants' insights is empathy-building information \footnote{\revision{It is essential to acknowledge the complexities of empathy in disability contexts ~\cite{Bennett_empathy}. Our primary goal is to facilitate effective help for BLV individuals. Thus, our focus is on the practical benefits of building empathy to serve the purpose of giving better help. We recognize this step towards empathy as a valuable, yet not central, aspect of our approach to help support. }} \revision{, serving as a means to the practical benefits of giving better help to the BLV participants.} 
We found that the information provided by help supporters has a positive impact on \revision{helping sighted participants to better understand how BLV participants perceive the environments with non-visual senses.}
%enhancing the empathy of sighted helpers towards their BLV helpees\revision{}
% It helped sighted helpers to understand and empathize with how BLV people use non-visual senses to perceive their surroundings, 
\revision{A}s a result, sighted helpers improved their performance during the collaboration phase. 

% Information that helpes sighted users to a mindset of empathy. Examples of this type of information include ``Don't point `over there' '', "Describe left and right from BLV person's perspective", "Use non-color descriptions", etc.

% For sighted helpers who had no prior experience interacting with BLV people, there are things dont come intuitively and they often forgot to say. We found that the messages provided by help supporters functions as empathy information, which remind them to empathzie with the needs of BLV individuals. 

We found that sighted helpers who did not have prior experience interacting with BLV individuals often lack awareness of visual concepts that need to be communicated verbally and often overlook the need to communicate them. 
Our findings revealed that the messages provided by help supporters serve as empathy information, acting as reminders for sighted helpers to view things from the BLV helpee's perspective.

% \begin{quote}
% \textit{``I think the corrections from the collaboration systems were helpful. Since most people don't have experience helping BLV people, we often forget various things important for BLV people, such as saying "goodbye" before leaving, using the walls as guidelines, or not using colors, for instance.''} -P2S
% \end{quote}

% Sighted users indicate that, after a few messages, they begin to empathize more with the BLV user’s senses which helped them get into a helping mindset. This is especially apparent when the sighted participant had prior experience helping BLV people sometime ago.

% Participant P4S indicated in the pre-study survey that they had prior experience helping a BLV classmate a few years ago. Seeing the messages provided by the system brought back a lot of her past memories and quickly helped her get into a mindset of thinking from BLV person's perspective. 

Sighted participants reported that, after a few messages from the help supporters, they developed a greater sense of empathy toward their BLV partners' non-visual sensory experience. They were able to adopt a more supportive mindset. 
This effect is particularly pronounced for sighted participants who had past experience interacting with BLV individuals, even if such experiences occurred some time ago.
% For instance, Participant P4S noted in the pre-study survey that they had previously helped a BLV classmate several years ago. The empathy information by \textit{help supporters} triggered a flood of memories related to these past interactions, facilitating a swift transition into a mindset that comprehends and embraces the needs of a BLV individual.

% \begin{quote}
%   \textit{``I have some experience helping from a long time ago with my friend in [home city]. But that was a long time ago. Like I’ve not been in that frame of mind for so long. So it was really helpful and nice to just remind myself. And, it makes you appreciate your senses. But also appreciate what other people are going through and appreciate that other people are navigating the world in a totally different way. And that we can still help each other despite our sensory differences.''} - P4S  
% \end{quote}

% \begin{quote}
%   \textit{``I have some experience helping from a long time ago with my friend in [home city]. I’ve not been in that frame of mind for so long. So it was really helpful and nice to just remind myself. It makes you appreciate what other people are going through and appreciate that other people are navigating the world in a totally different way. And that we can still help each other despite our sensory differences.''} - P4S  
% \end{quote}

\begin{quote}
  \textit{``I have \revision{[...]} experience helping \revision{[...]} my friend in [home city]. \revision{[...]} So it was really helpful and nice to just remind myself. It makes you appreciate what other people are going through and appreciate that other people are navigating the world in a totally different way. And that we can still help each other despite our sensory differences.''} - P4S  
\end{quote}

Furthermore, we found that empathy information from \textit{help supporters} reshaped sighted participants' thought processes during collaboration with their BLV partners, amplifying the empathy mindset into an educational dimension. 
Participant P9S described how one message taught them to shift their thinking to the BLV individual's perspective: 

\begin{quote}
\textit{ ``I also really liked one of the other hints from the [\textit{Pictorial Display}], where it said: `Are there any reference points that I can use?' That hint really changed my thinking. Because I was totally thinking `Just walk straight and then to the left.', but the hint made me think, `Okay, what can the BLV user use as a way to tell if he's going straight?'. And this made [giving the description] more challenging, but it also made it more useful.''} -P9S   
\end{quote}

%subsubsection{Validation Information}
\subsubsection{Validation: Sighted helpers want to know how well they are performing.}
\label{sec:finding_collab_validation}

% We found that sighted participants enjoyed having a sense of reassurance and validation while they help the BLV participants, letting them know that they are doing great and their collaboration is useful. 

% Some sighted participants reported that they preferred formal messages[\textit{Vague Directions Flagger}] because it is more straightforward and provides validation when they "checked the boxes". 

% We found that sighted participants often looked for a sense of reassurance and validation from help supporters throughout the course of their collaboration with BLV partners. They read into the signs from help supporters as information of their performance. For example, lack of transcript review highlights and the disappearance of one pictorial image to another can inform them they are doing great and they feel validated. 
The fourth type of information emerged from participants' experience is validation. 
We found that sighted participants frequently sought a sense of reassurance and validation from \textit{help supporters}. They interpreted cues from \textit{help supporters} as indicators of their performance. For instance, the absence of vague direction flags or the transition from one pictorial image to another was often interpreted as positive validation.

Participant P6S described how they elicited feelings of achievement by connecting the appearance and disappearance of pictorial messages and the perception of accomplishing subtasks during the collaboration: 

\begin{quote}
  \textit{“[\textit{Pictorial Display}] offered me a bit more reassurance. That's because it almost felt like when each image showed up and then went away, it was like some sort of task or mini-challenge that I was trying to tackle. Yeah, which was kind of strange, but like, somewhat gratifying."} - P6S
\end{quote}

Participant P9S provided their interpretation that the lack of highlights on \textit{Vague Directions Flagger} was a positive validation of their performance:

\begin{quote}
    \textit{``In [\textit{Vague Directions Flagger}], it was nice when the yellow highlighting showed up since it functioned as showing that okay, with no highlights, that means you're using good language, and you don't need to be corrected.''} -P9S
\end{quote}

% “I'm more confident when it comes to [\textit{Vague Directions Flagger}] since even I can see the transcription [reviews], so I can make corrections if necessary.” -P10S

% \subsubsection{Repetitive Information ~\\}

% We observed that information that nudges toward empathy is the type that gets annoying after sighted users successfully adopted the mindset and changed their behavior; while information asking for specificity may be worth repeating more frequently to solicit detailed information whenever only vague information is given. 

% Frequent messages are desired at the beginning of the helping session. As the participants learn and adopt the mindset, repetitive messages become annoying and unnecessary. 

While validation information is generally well-received by sighted participants, we also found that their positive effects can diminish and even become annoying once sighted helpers gain confidence in how to help BLV people well: 

%We observed that one specific message ``Describe direction from where [the BLV helpee] is facing'' aimed at encouraging sighted helpers to describe directions based on the BLV individual's orientation could provoke impatience and be perceived as redundant when the message is repeated with every directional instruction. 
%remind sighted people to describe the direction from where the BLV person is facing, and when the reminder occurs every time a direction is given, the sighted user expressed impatience and felt it was unnecessary. 

As participant P1S expressed: 
\begin{quote}
  \textit{``When I was giving directions and using the words `left', `right', etc., every single time [help supporters] would tell me to make sure that I was thinking from the perspective of the BLV person. This feedback got annoying quickly...''
  %[help supporters could] less frequently provide such a reminder, since I'm able to take that into account after the first time. But with regards to the other directions where I was not being specific enough, the systems were very helpful.''
  } -P1S  
\end{quote}

% \yy{Design Implication: \textit{help supporters} should provide real time feedback to sighted participants, validate what they did correctly, but the frequency of notifications should vary based on users' empathy mindset and performance. }

% \subsubsection{Command vs. Nudge ~\\}

\subsection{RQ4 Findings: Location of Information}
\label{sec:rq4-findings}

% The location of the information on the system is determined by the interdependent relationships between the sighted helper, the BLV person, and the environment. The relationship is fluid and dynamic over the course of a collaboration session. 

RQ4 is about identifying where help supporters should situate the information it provides during the collaboration phase. We evaluated two approaches represented by our prototypes. The \textit{Pictorial Display} positioned the information on the chest of the BLV helpee, making the information publicly displayed. The \textit{Vague Directions Flagger} involved situating the information on the personal smartphone of the sighted helper, thereby ensuring the information is privately displayed.
Through interviews and observations, we found that sighted participants valued maintaining eye contact with their BLV partners (Section 6.2.1). At times, when a sighted helper needs to assess the surroundings in order to convey the information to BLV helpee, they also expressed a preference for the information to be located near the environment (Section 6.2.2). Moreover, we found that maintaining the privacy of sighted participants' mistakes during their conversations is an important consideration (Section 6.2.3). 
% we found that, most of the time, sighted participants value eye contact with their BLV partner, and occasionally, when sighted helper needs to survey the environment they prefer the information to be adjacent to the environment as well. Additionally, sighted helpers value the privacy of their conversation.

% \begin{figure*}
%     \centering
%     \includegraphics[width=0.4\textwidth]{Figures/Location of Display.png}
%     \caption{
%     %Location of Information (Sketch - The dynamic environment should be depicted better.)
%     Sighted helpers' preference for where the AT's information should be located. We found that sighted people preferred the information to jump between locations to support two use cases: making eye contact with BLV helpees when talking to them (red), and looking at the envioronment to describe it (blue).
%     % The figure illustrates sighted helpers' preference for the location of the information. 
%     % When engaging in direct eye contact with BLV helpees, we found that sighted users exhibited a preference for positioning the information near the eyes or faces of BLV individuals. Conversely, when sighted helpers are surveying the surrounding environment, the optimal location for information is in proximity to the environment they are attempting to convey.
%     }
%     \label{fig:}
% \end{figure*}

\subsubsection{Sighted helpers want to maintain eye contact even with a supplemental display.}

We found that sighted participants valued eye contact with BLV participants.
%and the location of \textit{help supporters} should afford eye contact while providing information. 
Participants provided feedback on the location of both \textit{Pictorial Display} and \textit{Vague Directions Flagger} in the context of maintaining eye contact. 
For \textit{Pictorial Display}, some participants preferred the location of the display closer to the eyes, such as on a necklace or forehead, in order to maintain eye contact. 
In contrast, for the private display \textit{Vague Directions Flagger}, some participants expressed feeling uncomfortable checking their phone during a conversation.

% P4S highlighted the natural tendency to focus on their BLV partner's eyes during conversation and suggested that positioning the information display in close proximity to the eyes would prove the be more advantageous for upholding eye contact.

% \begin{quote}
% \textit{"Your natural reaction is to look at someone's eyes when you're talking to them...Yeah, mostly because it just kind of goes against your natural way of interacting with someone. Because you don't normally look at people and then look at their chest. If it was like a necklace or something, or it's like on your collarbone or something. Yeah, that could be useful...Maybe experimenting with different locations for it." } - P4S
% \end{quote}

% Sighted participants stated that eye contact is an important behavior in conversation. Although BLV person can't see the sighted helper, sighted helpers believe that it is still good practice to maintain eye contact.
P9S preferred \textit{Pictorial Display}'s location on the BLV individual, allowing sighted helpers to maintain eye contact with BLV helpees while engaging in conversation. 
Sighted participants believed that eye contact is important, even though due to BLV individual's situation, it might not be reciprocated in the traditional sense.
% This behavior aligns with the belief in the importance of eye contact, even in situations where it might not be reciprocated in the traditional sense.

\begin{quote}
\textit{``I also appreciated that the [\textit{Pictorial Display}] was on him [BLV helpee], so that you know, I could look at him and still talk to him at the same time, because I feel like I've been told that it's good practice to look at the person, even if they can't necessarily [see] you.'' } - P9S
\end{quote}

\subsubsection{Sighted helpers want to see the environment even with a supplemental display.}

% In addition to eye contact, we found that the location of information should also support sighted helpers to gather environmental information and formualte their descriptions when needed. 

% %In addition to locating the information in proximity to BLV individuals' eyes for sustaining eye contact, we found that there are instances 

% We found that sighted participants also need to have the information located in proximity to the environment in order for them to gather environmental information and formulate their descriptions to BLV helpees. 

Besides positioning the display close to BLV people's eyes, our findings revealed that sighted participants also want the information to be situated close to the environment. This arrangement enables them to learn the surrounding environment and give accurate descriptions to the BLV helpees.

% Collaboration Phase is dynamic. There are instances when sighted helper's attention shifts away from maintaining eye contact with BLV people, and focus on getting a better look at the environment such that they can provide an accurate description. 
% This scenario is specially common when the sighted helper is not familiar with the environment they are describing, they turn their attention to looking around and not always facing the BLV helpee during this time, and as a result, preferred that the information to be overlaying on the environment instead of the BLV helpee's person.

We observed that face-to-face helping has a dynamic nature. Sighted participants' attention transitions fluidly. There are instances when a sighted helper's focus shifts away from eye contact with BLV individuals, and towards gaining a better look at the environment to facilitate a more accurate description.
This phenomenon is particularly prevalent when the sighted helper is unfamiliar with the environment they are describing. They would momentarily break eye contact with the BLV helpee, and direct it towards the environment. As a result, these sighted participants preferred the information to be superimposed onto the environment instead.
%learning about the environment themselves before describing to the BLV participant. 
% Therefore, the sighted participant is not always facing the BLV participant as their position in the space shifts as they are trying to get a better look at the environment. 

P4S described the challenge they faced when shifting their view: 
%when attempting to view the information displayed on their BLV partner's chest, particularly when the environment plays a role in their collaboration, prompting them to shift focus from maintaining eye contact to observing the environment. P4S conveyed their challenge:

\begin{quote}
  \textit{``Describing the area and then looking at this little screen, so it's a bit kind of discombobulated.''} - P4S  
\end{quote}

% \begin{quote}
%     \textit{"A pro of the display is that you can easily see what the guidance or instruction that it's giving you via screen, as you're talking to [i.e. facing] the BLV user."} - P8S 
% \end{quote}

P3S recounted an instance when their physical positions with their BLV partner changed and the information located on the BLV partner's chest became occluded. 
%because the collaboration needed physical guidance, walking together while offering elbow. 
%As a result of shifting physical positions, 

\begin{quote}
\textit{``There's one situation as well, where he was walking next to me. He was holding my elbow or we were walking down the stairs, where I couldn't see the [\textit{Pictorial Display}] screen. You couldn't see the screen, so you couldn't really get the feedback.''}  –P3S
\end{quote}

In a dynamic environment, we found that some sighted participants preferred a static location of the display on their personal smartphones. 
The yellow highlights on \textit{Vague Directions Flagger} attract shifting attention and support glancing.

% Considering the dynamic nature of the Collaboration Phase, we found that some sighted participants preferred a static location of the information on their personal smartphones held in their hands. They described that, despite their attention shifting fluidly as they surveyed their environment, they found it convenient to periodically glance at the yellow highlights on their smartphone to access the information provided by \textit{help supporters}. 
% P4S and P5S detailed how they used peripheral attention to notice the yellow highlights: 

\begin{quote}
\textit{``I thought even just seeing the little highlighted yellow words from the periphery of my eye--that alone effectively served as a reminder to describe directions in a way that is going to be as clean as possible for the BLV user.''} -P4S
\end{quote}

% \begin{quote}
% \textit{``I think the [\textit{Vague Directions Flagger}] suggestions were more noticeable because they had highlighted areas. Whereas with [\textit{Pictorial Display}], I found myself looking around our environment and also looking at the BLV party’s eyes and their face more than the wearable device. So I would sometimes miss when there was a suggestion for a correction on directions.''} –P5S
% \end{quote}

\subsubsection{Sighted helpers want to keep their mistakes private.}

% We found that Location of information needs to consider privacy of the collaborative conversation. 
% When asked about their preference for public vs. private displays of the information, sighted participants valued the privacy of the conversation between BLV partner and themselves. 
% Some sighted people are self-conscious and embarrassed for their mistakes when interacting with BLV people, therefore, they prefer to receive corrective suggestions privately.
% Display publicly may amplify sighted people's learning mistakes while providing help, making sighted people self-conscious and embarrassed for their mistakes. 
% Therefore, private display [\textit{Vague Directions Flagger}] might be preferred for novice and less confident sighted helpers. 

Our findings showed that it is important to consider privacy when determining the location of information in the collaboration phase.
When we asked sighted participants about their preferences for public versus private displays, 
they conveyed that it is important to keep their conversation with BLV helpees private, especially their mistakes.
%they  importance of maintaining privacy of the conversation between themselves and their BLV partners, especially their mistakes in the conversation.
It emerged that some sighted participants experience self-consciousness and embarrassment when they make mistakes and need corrections from \textit{help supporters}. To make them more comfortable in the collaboration phase, they preferred to receive suggestive information discreetly on their personal smartphone device.

\begin{quote}
\textit{``I guess the first one [\textit{Vague Directions Flagger}] felt more personal, because other people can't see the corrections that are being made to me. I guess it would maybe depend on the person you know, maybe some people wouldn't want their mistakes to be on a screen [\textit{Pictorial Display}].'' }- P4S
\end{quote}

\section{Discussion}

Our study revealed many insights about how assistive technology can support face-to-face help between BLV and sighted strangers. In this section, we interpret those findings to pose design implications for help supporters and our community's other research efforts around interdependence and collaborative accessibility. 
The design implications and the findings from which they are derived are summarized in Table ~\ref{fig:table_designimplications}. 

% We organize our discussion in two sections. 
% Section ~\ref{sec:dis_in-situ} discusses the design implications drawn from the perspectives of both BLV and sighted participants for in-situ and real-time help support. 
% Our investigative approach is distinctive in that it considers the contextual involvement of all stakeholders, including the BLV helpee, the sighted helper, the \textit{help supporter}, the physical environment, and shared timing. This approach stands in contrast to prior works (refer to Section \ref{sec:related_collab}) which took a preplanned guideline approach or substituted sighted people's participation with assistive technology. 

% Section ~\ref{sec:dis_designforsighted} discusses the design implications for future \textit{help supporters} as technologies primarily serve sighted users during their interaction with BLV individuals. The research community increasingly embraces the social model of disability and includes people without disabilities as users of assistive technology (refer to Section ~\ref{sec:related_usedby}). 
% Our research makes a significant contribution to the underexplored sub-domain of assistive technology design, with a primary focus on design implications related to sighted users.
%our research makes a significant contribution to the underexplored domain of design of assistive technology primarily concerning sighted users. 

\begin{table*}
    \centering
    \caption{Summary of our design implications along with the finding(s) from which each is derived. The findings are summarized in Table ~\ref{fig:table_findings}, and the design implications are elaborated in the Discussion.
    }
    \includegraphics[width=0.8\textwidth]{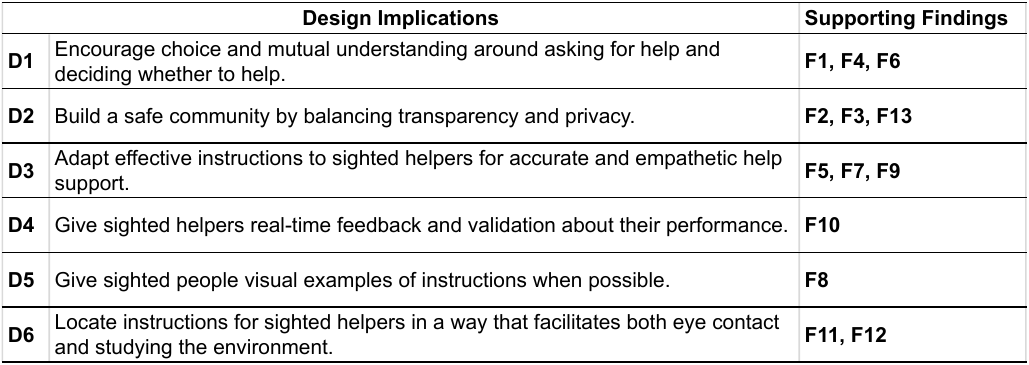}
    
    \label{fig:table_designimplications}
    \Description{Provides a summary of our design implications, along with the relevant finding(s) from which each is derived. The findings are those listed in Table 3, and the design implications are further elaborated upon in the Discussion.}
\end{table*}

\subsection{How Should Assistive Technology Support Co-located Help Between BLV and Sighted Strangers?}
\label{sec:dis_in-situ}

Our first two design implications (D1 and D2 in Table~\ref{fig:table_designimplications}) relate to how facilitating co-located help between BLV and sighted strangers.

\subsubsection{Encourage choice and mutual understanding around asking for help and deciding whether to help.}

In our study, we discovered common themes among both BLV and sighted participants, including the experience of social pressures, concerns about burdening the other party, and a shared desire to establish and navigate boundaries within the co-located help context. 
%BLV participants were concerned about the potential sighted helper's availability and wanted help from people who are located nearby and have free time to help (F4). 
\textit{Volunteer Platform} stood out as a preferred approach that reduces social pressures for both parties, largely owing to its ability to establish common ground and mutual understanding between them.
Connecting through an online platform before meeting in person provides the opportunity to communicate needs and expectations and establish mutual boundaries. 
Prior research made headways in understanding how social platforms facilitated abled people to connect online before shifting their interactions off the platforms and eventually meeting in person ~\cite{hsiao_people-nearby_2017}.  
Future research should investigate approaches that intersect both online and in-person social platforms, especially for mixed-ability populations, and explore designs that encourage choice and mutual understanding around asking for help and deciding whether to help. 

    % F1, The Volunteer Platform reduced social pressures for both parties.
    % F4, BLV people want help from people located nearby who have time.
    % F6, Both parties want to understand each other's intention toward help.

\subsubsection{Build a safe community by balancing transparency and privacy. }

When it comes to meeting strangers for help, our findings both align with and augment existing research.
Prior works found that 
BLV people are worried about personal safety in public and are hesitant to identify whom they ask for help from~\cite{thieme_i_2018, hurst_is_someone_there}. We discovered that sighted people share the same safety concerns. 
Synthesizing perspectives from both BLV and sighted participants, our design implication points to the importance of building a safe and welcoming community for all. The assistive technology itself can help establish that community through its platform.
The community should not only make sure that both parties feel safe to connect by facilitating mutual transparency of their identities but also create a non-judgmental and private space where sighted people can make and correct their mistakes.
We suggest that future work continue to explore the designs of in-person community-building technology by balancing transparency and privacy.

%D2, Build a safe community by balancing transparency and privacy. 

    % - F2 Both parties expect the platform to establish safe matches.
            
    % - F3 BLV people want their helpers to be BLV-friendly and, sometimes, female.
    %         transparency - exchange identity information 
    % - F13 Sighted helpers want to keep their mistakes private.
    %         privacy

% D5, Give sighted people visual examples of instructions when possible.
%     - F8 Sighted helpers want visual examples of help support guidelines.

\subsection{How Could Assistive Technology Facilitate Education for Sighted People?}
%help Supportrs used by sighted individuals to provide better help
\label{sec:dis_designforsighted}

Our remaining design implications (D3--D6) offer insights into how the accessibility community can shift the burden of education from BLV people (learning how collaborate with sighted people) to sighted people as well (learning how to collaborate with BLV people). By developing educational assistive technology for the many who would benefit from it, our community can further reframe assistive technology as not just being helpful for BLV people but rather for everyone.

%\subsubsection{Adapt instructions to sighted helpers for the particular BLV individual and deliver them at appropriate times. }
\subsubsection{Adapt effective instructions to sighted helpers for accurate and empathetic help support.}
%Adapt instructions to augment the sighted helper's knowledge with empathy and accuracy. 

%\yy{currently working on this paragraph}

%Due to the in-situ and real-time nature of help, there isn't a lot of time for reaction, 
Due to the in-situ and real-time nature of help, sighted helpers have limited time for immediate reactions to the help supporter's instructions, as observed in our study. 
Sighted helpers preferred help supporters to give them precise feedback on words and phrases requiring corrections.
Moreover, it is naturally easier for the sighted helpers to perform better without the need for repeated instructions when help supporters' instructions enabled them to empathize with BLV helpees' experience.
Future help supporter designs should adapt effective instructions to sighted helpers that empower them to make instantaneous improvements. 
\revision{Some prior work has used AI in learning technologies to tailor learning content to meet students' level of understanding in real-time ~\cite{molenaar_personalisation_2021}.}
% Our research underscores specific and empathy-building information as key factors in constructing effective instructions. 
Further research is needed, however, to explore the at-scale generation of instructions in real-time \revision{mixed-ability scenarios}. 
%explore the construction and deployment of such instructions on a larger scale in real-time scenarios

    % F5, BLV people want help from people knowledgeable about how to help.
    % % F7, Sighted helpers want to know how to correct their descriptions.
    % F7, Sighted helpers want to know the precise corrections they need to make. 
    % F9, Sighted helpers want to empathize with BLV people's experience.

\subsubsection{Give sighted helpers real-time feedback and validation about their performance. }

In our study, even though our prototypes did not explicitly include a validation feedback feature, it was emergent that sighted participants often sought reassurance from help supporters to gauge their performance and gain confidence (see Section ~\ref{sec:finding_collab_validation}). 
% Help supporters played a role in signaling when sighted participants were on the right track. Future design should explore how to combine correction and validation in real-time feedback and provide the information to sighted helpers when they need it. Contextual adaptation might be considered to provide the right amount of feedback without burdening or distracting sighted helpers. 
\revision{Previous work has found that in a human-to-human context, helpers of people with disabilities benefit from feedback and validation, these findings can be translated to help supporters, highlighting the universal value of feedback in supportive interactions ~\cite{anderson2011learning}. Additionally, since the way people interact with feedback changes depending on their stage of learning, it is imperative that the feedback itself evolves accordingly, ensuring its continued effectiveness and relevance ~\cite{hattie2016learning}. }
\revision{This highlights the need for p}ersonalization might be considered to match with sighted helpers' level of confidence and need for assurance. 

    % - F10 Sighted helpers want to know how well they are performing.

\subsubsection{Give sighted helpers visual examples of instructions when possible. }
    %- F8 Sighted helpers want visual examples of help support guidelines.

In the \textit{Pictorial Display}, sighted participants valued how easily they could look at the visual examples demonstrated by the Bitmoji characters and understand the right way to help (e.g. offering an elbow for guidance, following an edge to walk straight). Future help supporter designs can explore diverse ways of displaying visual examples to sighted helpers. 
For example, drawing upon previous research that employed AR to provide manufacturing workers with 3D instructions ~\cite{henderson2011armar}, sighted helpers might benefit from holographic visual demonstrations. These demonstrations could guide them on how to interact behaviorally with BLV people, encouraging actions that are helpful and avoiding actions that might startle or put them at risk.
 
% /AR manufacturing examples?

\subsubsection{Locate instructions for sighted helpers in a way that facilitates both eye contact and studying the environment.}
    % F11, Sighted helpers want to maintain eye contact even with a supplemental display.
    % F12, Sighted helpers want to see the environment even with a supplemental display.

We found that face-to-face help requires sighted helpers 
to navigate their visual attention between various people and places. In our study, sighted helpers wanted to maintain eye contact with BLV helpees and keep the environment in their sight in order to give accurate descriptions, all while simultaneously ensuring they are catching instructions provided by the help supporter. 
Future help supporter designs should consider locating instructions for sighted helpers in a manner that aligns with the dynamic nature of their attention. This could involve leveraging eye-tracking and AR technologies to display the instructions always within the sight helper's field of view. 
Furthermore, \revision{drawing from prior work in designing multimodal feedback for multitasking ~\cite{kim2011designing}, future designs could incorporate} multimodal feedback mechanisms \revision{to} provide sighted helpers with multiple streams of information through the appropriate modality and guide their attention toward the relevant people or places. 

\subsection{\revision{Demonstrating the Interdependence Framework in Mixed-Ability Scenarios}}
\label{sec:interdependence}

\begin{figure*}
    \centering
    \includegraphics[width=1.0\textwidth]{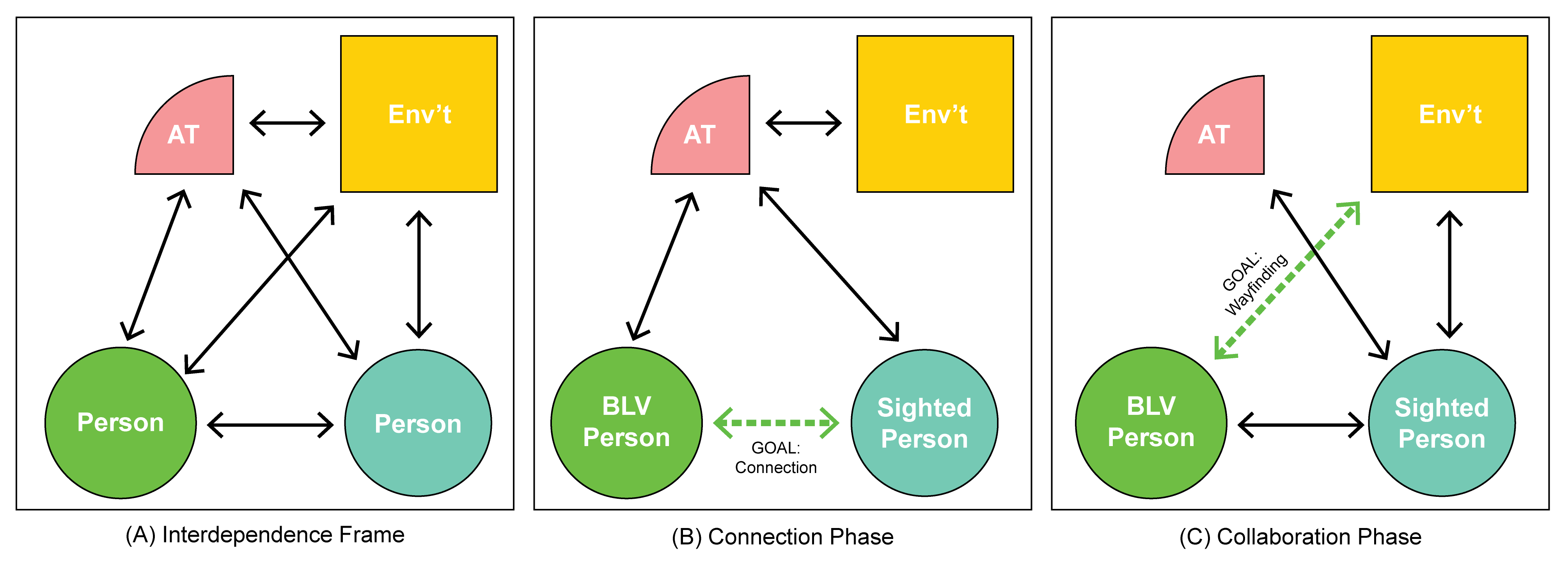}
    \caption{\revision{ An illustration of how our approaches demonstrate and explore the interdependence framework. (A) shows a reproduction of the original interdependence framework ~\cite{bennett_interdependence_2018}; 
    (B) shows, in the Connection Phase, assistive technology interacts with both BLV and sighted strangers to meet for help; 
    (C) shows, in the Collaboration Phase,  assistive technology provides the sighted helper with instructions such that they can best convey information about the environment to the BLV person.
}
    }
    \label{fig:interdependence}
    \Description{An illustration of how our approaches articulated the interdependence framework and expanded to include both BLV and sighted people. Three graphs are shown side by side. The leftmost graph is labeled “(A) Interdependence Frame” and has four nodes labeled “Person,” “Person,” “AT,” and “Environment,” with bidirectional edges fully connecting all nodes. The middle graph is labeled “(B) Connection Phase” and has four nodes labeled “BLV Person,” “Sighted Person,” “AT,” and “Environment.” In this graph, a dotted bidirectional edge connects “BLV Person” and “Sighted Person” and has the label “GOAL: Connection.” A bidirectional edge connects the “AT” node to every other node, but no other edges exist. The rightmost graph is labeled “(C) Collaboration Phase” and has four nodes labeled “BLV Person,” “Sighted Person,” “AT,” and “Environment” similar to the middle graph. In this graph, a dotted bidirectional edge connects “BLV Person” and “Environment” and has the label “GOAL: Wayfinding.” A bidirectional edge connects the “Sighted Person” node to every other node, but no other edges exist.}
\end{figure*}

\revision{
The concept of help supporters and its two constituent phases allowed us to demonstrate and explore how assistive technology can respect the interdependence framework proposed by Bennett et al.~\cite{bennett_interdependence_2018} rather than attempting to ``bridge a perceived gap between disabled bodies and environments designed for non-disabled people''~\cite{bennett_interdependence_2018}. The interdependence framework (Figure~\ref{fig:interdependence}(A)) argues that the goal of making the world accessible is not one that assistive technology should by itself bridge for a disabled person, but rather is a shared goal that people with disabilities, technology, surrounding people, and environmental infrastructure collaborate towards achieving. The framework thus suggested exciting new designs for assistive technologies that interact not only with disabled people but also with the environment and other people around them. Our research-through-design process for help supporters allowed us to test different interdependence configurations and discover users' attitudes and behaviors toward them.
}

\revision{
Figure~\ref{fig:interdependence}(B) illustrates the interdependence relations that we explore in the connection phase. Recall that, in the connection phase, the goal is to connect a BLV person and sighted person together to start collaborating---that is, to form the dotted edge at the bottom of Figure~\ref{fig:interdependence}(B). An assistive technology that aims to do this plays as a central liaison between the BLV person, each surrounding sighted person, and the environment---for example, using its knowledge of the environment to remove sighted people that are far away from consideration. Both connection phase prototypes follow this model but in very different ways, with the Person-Finder Glasses being situated very close to the BLV person and the Volunteer Platform being a more equidistant liaison. Our findings from these prototypes (F1--F6 in Table~\ref{fig:table_findings}) reveal the implications of this arrangement.
}

\revision{
Figure~\ref{fig:interdependence} (C) illustrates the interdependence relations that we explore in the collaboration phase. Recall that the goal for the collaboration phase is to give BLV people the information needed to find their way in the environment---that is, to form the dotted edge in Figure~\ref{fig:interdependence} (C). Now, it is the sighted person that acts as a liaison between the BLV person and the environment, but because sighted people often struggle to provide effective help for BLV people, the assistive technology works to not only advocate for the BLV person's needs, but also educate the sighted person on how to properly interact with and convey information to the BLV person. Both the Pictorial Display and Vague Directions Flagger follow this model but in very different ways. Our findings suggested that BLV people prefer knowledgable helpers and find that this role for assistive technology alleviates their burden to explain themselves (Finding F5), while sighted people find real-time feedback (Design Implication D4) and visual examples along with instructions (D5) served them well.
}

\subsection{\revision{Integrating with BLV People’s Current Practices}}

\revision{Recall from Section ~\ref{sec:related_currentpractices} that BLV people navigate and make sense of their surroundings using a unique set of current practices. 
In Thieme et al.’s review of BLV people’s current practices ~\cite{thieme_i_2018}, the authors identify opportunities for assistive technology to enable BLV people to better locate others around them and choose whom to interact with, foster a shared understanding of other people’s actions, and consider existing social relationships.}
\revision{Our findings around users’ attitudes and preferences for assistive technology respond to these visions, filling in some of the social gaps that the scenario in illustrates.}

\revision{Finding F1, for example, reveals that the design of a volunteer platform connecting BLV individuals and sighted strangers reduces social pressure for both parties when compared to BLV people approaching sighted individuals directly for help. 
Nevertheless, making a request through a volunteer platform might require more and effort than approaching a stranger directly.
Additionally, BLV people who are comfortable with approaching others using current practices may choose to do so, and assistive technology should not hinder BLV people’s autonomy in choosing how they navigate socially.}

\revision{Regarding the collaboration phase, we found that sighted people would like to learn more about BLV people’s experiences and processes (Finding F9) and their own current performance in giving help (Finding F10).
The burden of both tasks usually falls to the BLV person, who may be at times excited to offer feedback but under other circumstances prefer not to do so, this could depend on the situation and their amount of free time.
Our findings underscore the importance of supporting the social connections that BLV people wish to establish within their navigational process, which means that the approaches that we explored for the collaboration phase (Pictorial Display and Vague Directions Flagger) should only be used in cases where the BLV person does not want to do all of the advocacy and education for sighted people themselves. In Appendix~\ref{sec:autonomy-findings}, we show that the collaboration phase prototypes may take conversational agency
away from BLV individuals. Future work should explore collaboration phase designs that let the BLV person determine when the assistive technology should remain inactive and when it should intervene to facilitate the education of the sighted person.
}

%%%% 7_discussion.tex ends here %%%%

%%%% 8_limitations.tex starts here %%%%

\section{Limitations}

% 1. no field study

% we did not do a field deployment, some findings are speculative.
% some of the insights were based on 

Our research is limited in that participants only had a single encounter with the prototypes in designated study environments. While we were able to draw meaningful insights from their experiences, a field deployment with a longitudinal study would have allowed us to conduct observations of usage in a more diverse range of physical environments and use cases, and gain an understanding of how people would engage with the prototypes over time. \revision{This would shed light on how a mixed-ability help-based community forms and evolves, as well as how the information needed from the help supporters might change over time.}
Additionally, our participants only included people who live in a major city in the United States, and our sighted participants only included university students. Their preferences might not be representative of the BLV and sighted communities at large. \revision{University students may be more receptive to the help supporters' instructions than the average public. Future studies with helper populations from other education backgrounds may uncover additional design implications on their preferences for the formats of instructions.}
Furthermore, we only \revision{studied} four help supporter prototypes. While we \revision{selected designs that covered very different points in the design space for both the connection and collaboration phases}, there could exist other types of approaches we did not explore and experiment with. 

% 4. the approaches represented by the prototypes might not have comprehensive coverage.
% There are other types of approaches we did not address.
% Future work should investigate them

%%%% 8_limitations.tex ends here %%%%

%%%% 9_conclusion.tex starts here %%%%

\section{Conclusion}

In this research, we explored the design space of assistive technology to support face-to-face help between BLV and sighted strangers.
We proposed \revision{diverse approaches toward} assistive technology \revision{serve as} \textit{help supporters}, which collaborates with both parties throughout the help process. 
We evaluated four approaches spanning two phases: the connection phase (finding someone to help) and the collaboration phase (facilitating help after finding someone). 
These approaches are represented by \revision{design prototypes}: 
\textit{Person-Finder Glasses} and \textit{Volunteer Platform} for the connection phase, \textit{Pictorial Display} and \textit{Vague Descriptions Flagger} for the collaboration phase. 
Our findings from a 20-participant mixed-ability study reveal how the help supporters can best facilitate connection, which types of information they should present during both phases, how the display should be situated with the environment, and more. 
% We posed six design implications that guide future directions of designing assistive technology for co-located mixed-abilities help, and educating sighted people
% of designing assistive technology for mixed-abilities community building and educating sighted people.
% We discuss design implications for future approaches to support help.
% , and hope that shift the burden of learning how to collaborate with others from BLV people alone to sighted people as well
Our design implications reveal future directions for assistive technology that fosters mixed-ability help and for shifting the burden of education toward sighted people.

% Blind and low-vision (BLV) people face many challenges when venturing in public environments, often wishing it were easier to get help from people nearby. Ironically, although sighted people would often be happy to help, help rarely happens. 
% Asking for help is socially awkward for BLV people, and sighted people lack experience helping BLV people. In this work, we explore a new category of assistive technology called \textit{Help Supporters}, which collaborate with both parties throughout the help process. We explore the design space of help supporters by investigating four diverse approaches for supporting help. These span two phases: the connection phase (finding someone to help) and collaboration phase (facilitating help after finding someone). Our findings from a 20-participant mixed-ability study reveal how the help supporters can best facilitate connection, which types of information they should present during both phases, and more. We discuss design implications for future approaches to support help.

% \section{Acknowledgement}
% Michael Malcolm

%%%% 9_conclusion.tex ends here %%%%

\begin{acks}
\minor{We thank Michael Malcolm, an individual with low vision, for participating in our mixed-ability co-design process. Michael Malcolm was a summer intern in our lab working on another project.}
\end{acks}

\bibliographystyle{ACM-Reference-Format}
\bibliography{source/chi24-920}

\appendix

%%%% 10_appendix.tex starts here %%%%

%TC:ignore

\newpage
\section{Additional Findings: How Help Supporters Affect BLV Users' Feelings of Autonomy}
\label{sec:autonomy-findings}

% \subsubsection{Preserving BLV Conversational Agency ~\\}

%Both the \textit{Pictorial Display} and \textit{Vague Directions Flagger} advocate for the BLV person and help the sighted person give effective help, but doing so runs the risk of burying the BLV person's own voice and autonomy in the process of help. 
While both the \textit{Pictorial Display} and \textit{Vague Directions Flagger} supported the sighted helpers in giving effective help, they also have an effect of advocating for the BLV person.
This presents a risk of burying the BLV person's own voice and autonomy in the process of help. 
Both BLV and sighted participants raised several such concerns, with the general consensus that, if implemented poorly, collaboration systems may take conversational agency away from BLV individuals:
% We had several interesting points of feedback from our participants Here we report on our findings 
% Our Collaboration Phase prototypes also revealed how BLV participants felt about the help supporters 
% ' feeling toward  Here we report our findings on participants’ concerns about help supporters’ corrective suggestions replacing some of the BLV individual’s agency in asking clarification questions. 

\begin{quote}
\textit{“In general, overall, I think I prefer asking clarifying questions myself rather than having the system ask them.”} -P5B
\end{quote}

\begin{quote}
\textit{
“The best feedback I can think of would be [from] the BLV person themselves. For instance, when they say, “Can you explain more?” or when they ask follow-up questions."} -P3S
\end{quote}

% In short, participants’ general consensus was a concern that, if implemented poorly, collaboration systems may disenfranchise BLVs by taking conversational agency away from them. 

%Participants also shared a concern that help supporters could feel dehumanizing or marginalizing to the BLV person if the help supporter became the primary guide for enforcing social etiquette and effective instructions.
Participants also shared a concern that using help supporters too much could feel insensitive to the BLV person if the help supporter became the primary guide for enforcing social etiquette and effective instructions.

\begin{quote}
\textit{
“[The helper] reading the \textit{Pictorial Display} or the \textit{Vague Directions Flagger} while I'm talking to them [...] felt a little bit strange. It’s a little like, ‘Oh, they have to read the instruction manual for how to interact with a person like you.’"} - P8B
\end{quote}

However, participants also mentioned the value and importance of having collaboration systems, as well as vital reminders they provide to sighted helpers:

\begin{quote}
\textit{
“Maybe there's a cliff behind me, and I don't know it. So, the help supporter is really important, because it can remind the sighted person, "The BLV needs to know this." But you have to be careful to make sure that it is telling the essential things, and leaving other things up to the BLV individual to say.”} -P2B
\end{quote}

These sentiments reveal a design tension between nudging sighted helpers toward giving better directions and maintaining organic communication between the BLV and sighted individuals. In the balancing act that help supporters must navigate between the two extremes, one BLV participant highlighted the importance of preserving BLV individuals’ opportunity to speak up and advocate for themselves:

\begin{quote}
\textit{
“One of the huge problems in the blind community is that blind people oftentimes won’t express their needs...I think there’s a balance that has to be achieved here. You want to give the sighted world information it needs [to help BLV people] without taking advocacy away from the blind person. [...] With the corrections that the collaboration systems automatically give to the sighted helper, those may take away some of the BLV individual’s agency.”} -P2B
\end{quote}

One way of achieving this balance may be for the help supporter to let the BLV person control whether they would like to lead the sighted person's guidance or let the help supporter do the heavy lifting.

\section{Messages used for the Pictorial Display Prototype}
\label{sec:bitmoji-reference}

The set of 20 pictorial display messages are designed by the authors using Bitmoji ~\cite{bitmoji_nodate} as a source of the avatars. 

\includegraphics[width=0.38\textwidth]{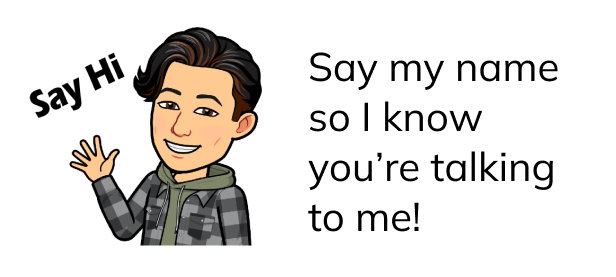}

\includegraphics[width=0.38\textwidth]{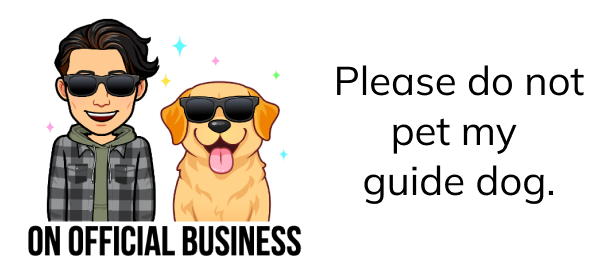}

\includegraphics[width=0.38\textwidth]{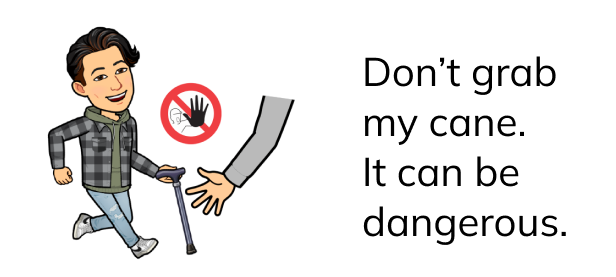}

\includegraphics[width=0.38\textwidth]{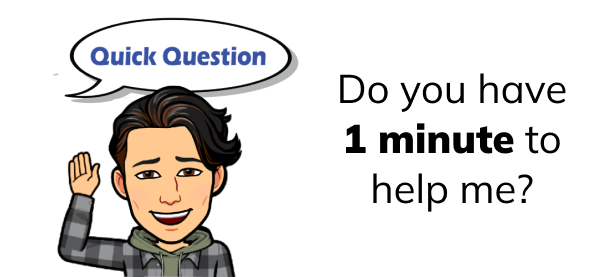}

\includegraphics[width=0.38\textwidth]{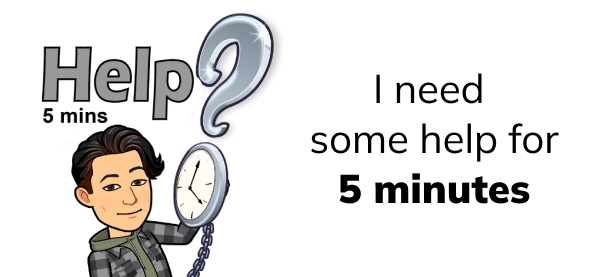}

\includegraphics[width=0.38\textwidth]{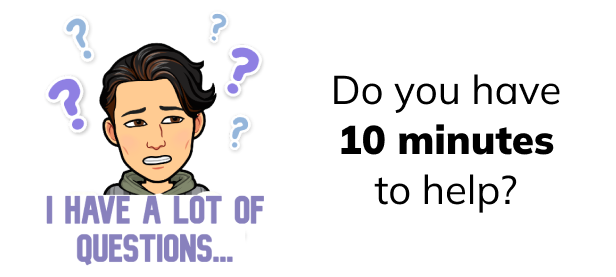}

\includegraphics[width=0.38\textwidth]{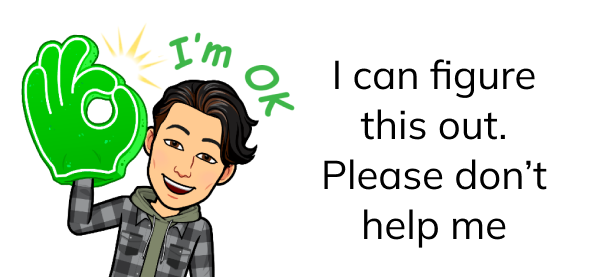}

\includegraphics[width=0.38\textwidth]{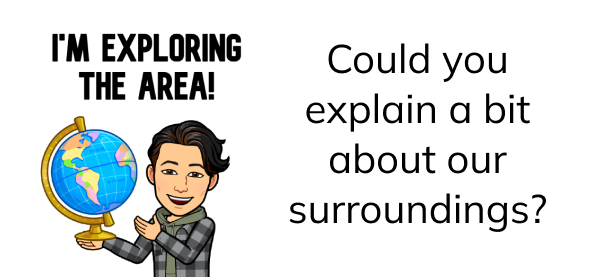}

\includegraphics[width=0.38\textwidth]{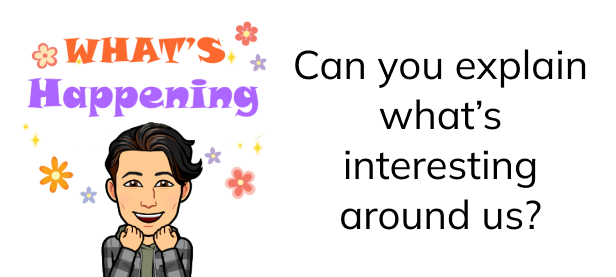}

\includegraphics[width=0.38\textwidth]{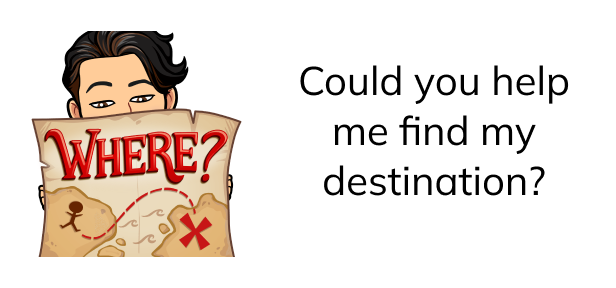}

\includegraphics[width=0.38\textwidth]{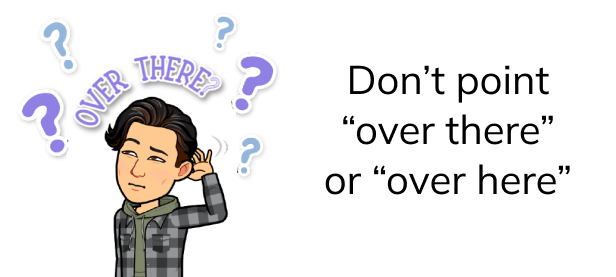}

\includegraphics[width=0.38\textwidth]{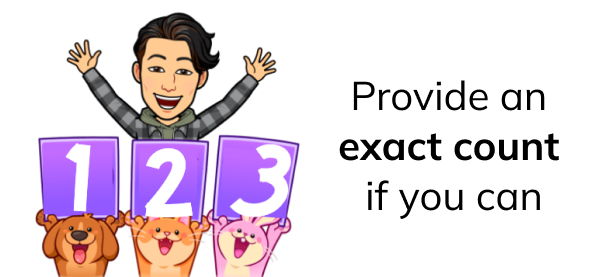}

\includegraphics[width=0.38\textwidth]{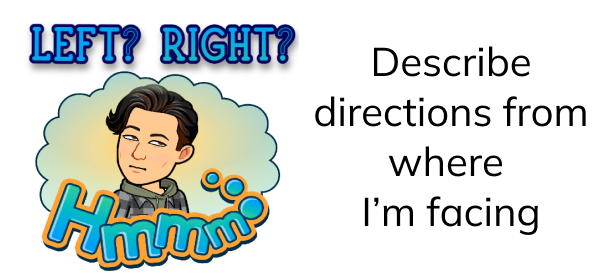}

\includegraphics[width=0.38\textwidth]{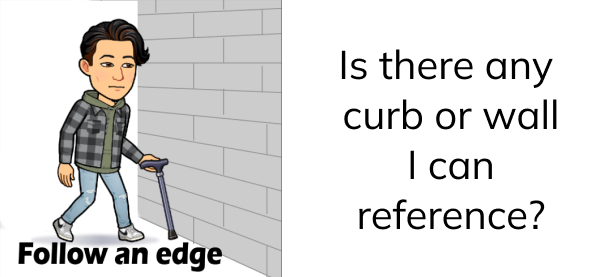}

\includegraphics[width=0.38\textwidth]{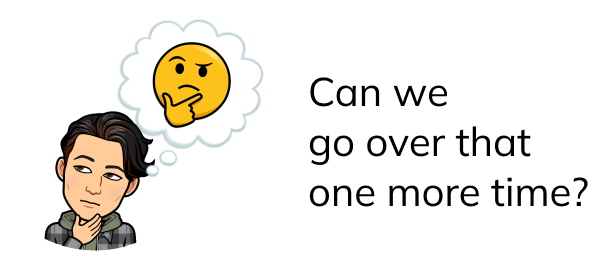}

\includegraphics[width=0.38\textwidth]{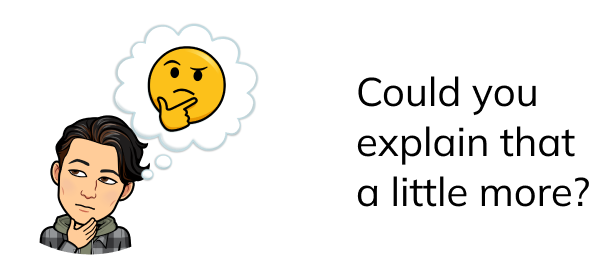}

\includegraphics[width=0.38\textwidth]{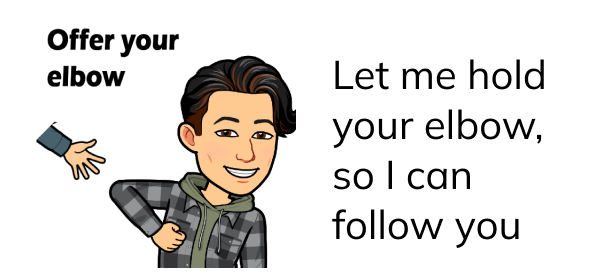}

\includegraphics[width=0.38\textwidth]{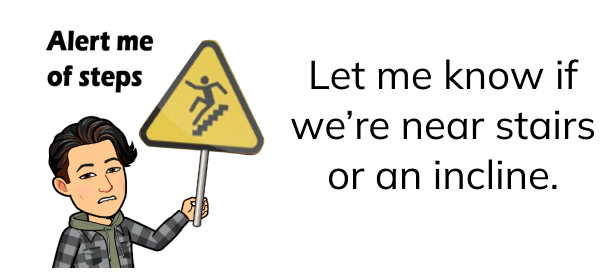}

\includegraphics[width=0.38\textwidth]{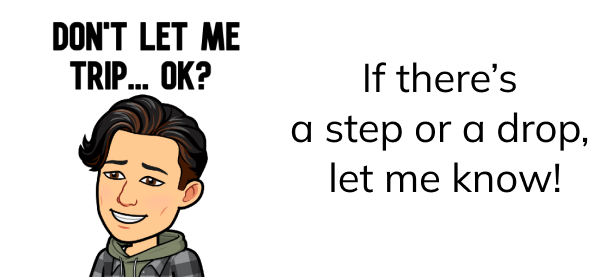}

\includegraphics[width=0.38\textwidth]{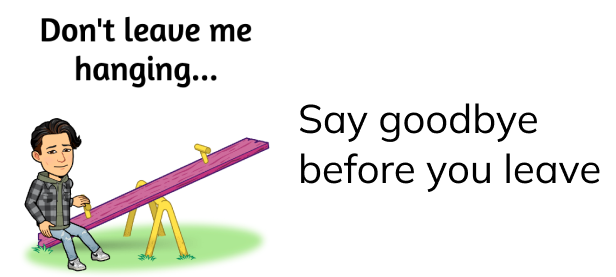}

\twocolumn
\section{Rules of Vague Directions Flagger}
\includegraphics*[width=0.7\textwidth]{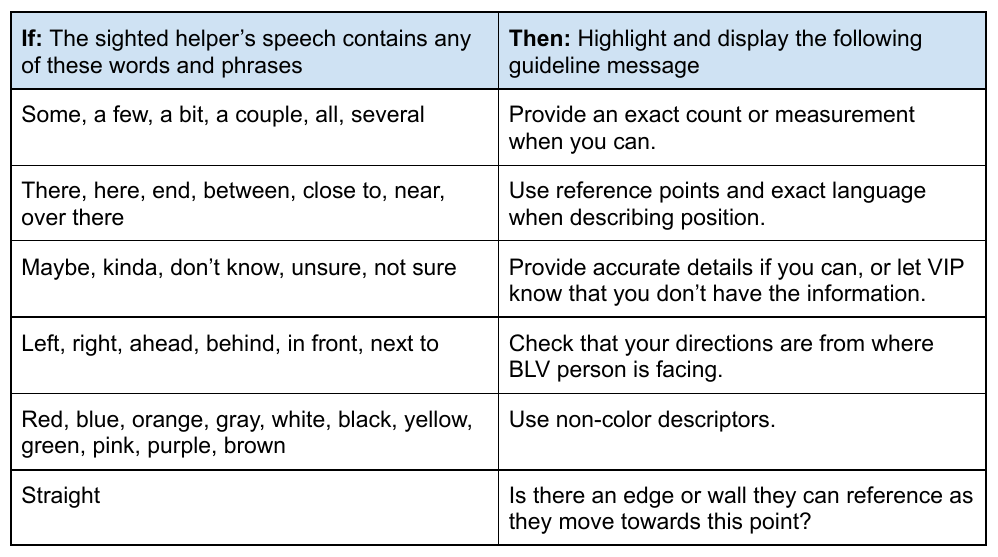}
\label{sec:rules_vaguedirectionsflagger}

%TC:endignore

%%%% 10_appendix.tex ends here %%%%

\end{document}